\documentclass[pdftex,twocolumn,epjc3]{svjour3}          

\RequirePackage[T1]{fontenc}

\smartqed  

\RequirePackage{graphicx}
\RequirePackage{mathptmx}      
\RequirePackage{flushend}
\RequirePackage[numbers,sort&compress]{natbib}
\RequirePackage[colorlinks,citecolor=blue,urlcolor=black,linkcolor=blue]{hyperref}
\usepackage{amssymb,amsmath}
\makeatletter
\newcommand\footnoteref[1]{\protected@xdef\@thefnmark{\ref{#1}}\@footnotemark}
\makeatother

\journalname{Eur. Phys. J. C}

\begin{document}

\title{Multiloop contributions to the on-shell-$ \overline{\rm{MS}}$ heavy quark mass relation in QCD and the asymptotic structure of the 
corresponding series: the updated consideration
}

\author{A. L. Kataev
 \thanksref{e1,addr1
 }
        \and
        V. S. Molokoedov 
\thanksref{e2,addr1,addr2} 
}

\thankstext{e1}{e-mail: kataev@ms2.inr.ac.ru}
\thankstext{e2}{e-mail: viktor\_molokoedov@mail.ru}

\institute{Institute for Nuclear
Research of the Russian Academy of Sciences, 60th October Anniversary prospect 7a, Moscow, 117312, Russia\label{addr1}
          \and
          Moscow Institute of Physics and Technology, Institusky per. 9, Dolgoprudny, Moscow region, 141700, Russia\label{addr2}
}

\date{Received: date / Accepted: date}

\maketitle

\begin{abstract}
The asymptotic structure of the QCD perturbative relation between the on-shell and $\overline{\rm{MS}}$  heavy quark masses is studied. We estimate the five and six-loop contributions to this relation by three different techniques. First, the effective charges motivated approach in two variants is used. Second, the results following from the large-$\beta_0$ approximation are analyzed.
Finally, the consequences of applying the asymptotic  renormalon-based formula are investigated. We show that all  approaches lead to corrections which are qualitatively consistent in order of magnitude. Their sign-alternating character in powers of the number of massless quarks is demonstrated. We emphasize that there is no contradiction in the behavior of the fine structure of the renormalon-based estimates with other approaches if one use the detailed information about the normalization factor included in the renormalon asymptotic formula. The obtained five- and six-loop estimates indicate that in the case of the $b$-quark the asymptotic character of the studied relation manifests itself above the fourth order of PT, whereas for the $t$-quark it starts to reveal itself after the seventh order. This allows to conclude that like the running masses, the pole masses of the $b$ and especially $t$-quark
in principle may be used in the phenomeno- logically-oriented studies.
\end{abstract}

\section{Introduction}

It is well known that the bare  masses of 
quarks in QCD can be expressed through their renormalized finite analogs defined  in a particular scheme. In this work we will consider primarily two renormalization schemes, namely the $\overline{\rm{MS}}$- and the on-shell 
$\rm{OS}$-scheme. The latter is used for defining the pole masses of heavy quarks. The relevant renormalization prescriptions for masses of these particles have the following form:
\begin{eqnarray}
\label{defOS}
m_{0,q}=Z^{\rm{OS}}_mM_q, ~~~
m_{0,q}=Z^{\overline{\rm{MS}}}_m\overline{m}_q, 
\end{eqnarray}
where $m_{0,q}$, $M_q$ and  
$\overline{m}_q$  are the bare, pole and $\overline{\rm{MS}}$-scheme running  masses respectively. The renormalization mass constants 
$Z^{\rm{OS}}_m$ and  $Z^{\overline{\rm{MS}}}_m$ contain the traces of ultraviolet divergences in the form of poles and are represented by the perturbation theory (PT) series  in powers of the coupling constant of the strong interaction depending on the scale parameter $\mu$ and defined in the corresponding subtraction scheme.

Due to the fact that the masses $M_q$ and 
$\overline{m}_q(\mu^2)$ are the finite renormalized  quantities,  
their ratio must also be finite. 
It is convenient to introduce the following  relation between the pole and  running  masses of 
heavy quarks, also  called in the literature as the 
on-shell-$\overline{\rm{MS}}$ mass relation:
\begin{equation}
\label{expansion}
\frac{M_q}{\overline{m}_q(\mu^2)}=1+\sum\limits_{k=1}^{\infty} \tilde{t}^{\;M}_k(n_l, n_h, \mu^2/M^2_q) a^k_s(n_f, \mu^2)
\end{equation}
with the strong coupling constant $a_s=\alpha_s/\pi$ defined  in the $\overline{\rm{MS}}$-scheme
in the Minkowski time-like region. The number of the active flavors $n_f$ running inside the fermion loops (the values $n_f=4,5,6$ correspond to the cases of the charm, bottom and top-quarks respectively) is related
 to the number of the light (massless) $n_l$ and heavy (massive) $n_h$ flavors by the following way $n_f=n_l+n_h$. In this work we use the approximation when only one heavy quark is massive i.e. $n_h=1$ 
 and the rest $n_l=n_f-1$ are massless.

The one-loop term $\tilde{t}^{\;M}_1$ was calculated a long time ago in \cite{Tarrach:1980up}. The two-loop correction $\tilde{t}^{\;M}_2$ was analytically  computed in \cite{Gray:1990yh} and confirmed later in \cite{Avdeev:1997sz, Fleischer:1998dw}. The $\mathcal{O}(a^3_s)$ contribution was evaluated independently by  analytical  \cite{Melnikov:2000qh} and semi-analytical  \cite{Chetyrkin:1999qi} methods. 

In the relation (\ref{expansion}) for coefficients $\tilde{t}^{\;M}_k(\mu^2/M^2_q)$ the transition from the pole mass to the running one can be carried out by solving the corresponding renormgroup (RG) equations. After this, it is possible to define the coefficients $t^M_k(\mu^2/\overline{m}_q(\mu^2))$. In the normalization point $\mu^2=\overline{m}^2_q$ we have $t^M_k((\mu^2=\overline{m}^2_q)/\overline{m}^2_q)=t^M_k(1)\equiv t^M_k$ and
\begin{equation}
\label{t^M}
\frac{M_q}{\overline{m}_q(\overline{m}^2_q)}=1+\sum\limits_{k=1}^{\infty} t^M_k a^k_s(\overline{m}^2_q).
\end{equation}

In any order of PT the terms $t^M_k$ can be expanded in powers of the number of $n_l$ and $n_h$. Fixing $n_h=1$ we arrive to the following expansion:
\begin{equation}
\label{zim}
t^{M}_k=\sum\limits_{i=0}^{k-1} t^M_{k, i} n^i_l.
\end{equation} 

In particular, the four-loop coefficient $t^M_4$ is a third degree polynomial in $n_l$:
\begin{equation}
\label{zm4}
t^{M}_4=t^M_{4, 3}n^3_l +t^{M}_{4, 2}n^2_l +t^{M}_{4, 1}n_l+  t^{M}_{4, 0}~.
\end{equation}

The first two coefficients $t^M_{4, 3}$ and $t^M_{4, 2}$  in (\ref{zm4}) were calculated analytically in \cite{Lee:2013sx}. Note that the exact numerical expressions of the  terms leading in powers of $n_l$ in (\ref{zim}) were obtained in \cite{Ball:1995ni} up to the ninth order of PT from consideration of the contributions generated by a renormalon-type chain of quark loops inserted into the gluon line which renormalizes the propagator of the heavy massive quark. The last two coefficients  in (\ref{zm4}) have not yet been computed in the analytical form. However, after the numerical evaluations performed in \cite{Marquard:2015qpa} for the overall term $t^M_4$ at the fixed number $n_l=3,4,5$, 
the approximate values of the contributions $t^M_{4,1}$, $t^M_{4,0}$ have been obtained with the help of the least squares method (LSM) in \cite{Kataev:2015gvt}. It allows to solve the overdetermined systems of algebraic equations and also to fix the uncertainties of its solutions. The similar expressions for these two terms were also found in \cite{Kiyo:2015ooa} by means of a special fitting procedure. It was based on the application of the renormalon calculus of 
\cite{Beneke:1994qe, Bigi:1994em, Beneke:1994sw, Beneke:1994rs, Beneke:1998ui} and more definitely on the renormalon asymptotic formula for coefficients of the relation between the pole and running masses of heavy quarks originally derived in \cite{Beneke:1994rs, Beneke:1998ui}. 
 
Later on the evaluation of the $t^M_4$-coefficient was done in \cite{Marquard:2016dcn} with higher precision than in \cite{Marquard:2015qpa} and for a much larger number of  flavors in the range $0\leq n_l \leq 20$. The central values of terms $t^M_{4, 0}$ and $t^M_{4, 1}$, being also extracted in \cite{Marquard:2016dcn} from the fitting of the numerical results for $t^M_4$ at the fixed number of $n_l$, 
 agree with the ones obtained with the help of the LSM and presented in the ``Note added'' of \cite{Kataev:2015gvt} and in \cite{Kataev:2018sjv}. 
 
Despite the apparent smallness of the four-loop corrections  to the on-shell-$\overline{\rm{MS}}$ heavy quark mass relation, its knowledge is important from both phenomenological and theoretical points of view.  
Indeed, the asymptotic nature of the PT series\footnote{See a well-known Dyson's pioneering work \cite{Dyson:1952tj} on this topic.} for the relation between the pole and running masses, which is governed by the dominant $u=1/2$ infrared renormalon contributions to the Borel image of this relation (discovered in \cite{Bigi:1994em, Beneke:1994sw}), leads to the factorial growth of the coefficients $t^M_k$ at the large orders $k$. This fast increase is associated with the  sensitivity of the pole mass to small momenta, due to which it suffers from the large perturbative corrections \cite{Beneke:1994rs, Beneke:1994sw}. On the contrary, the running mass, defined within the $\overline{\rm{MS}}$-scheme, depends on the ultraviolet (UV) subtraction of divergences only and, therefore, does not contain the infrared (IR) renormalon contributions. In this regard, it is very important to know in the specific physical studies when the asymptotic behavior will begin to manifest itself in the definite cases of the charm, bottom and top quarks.

As follows from the results of \cite{Gray:1990yh, Avdeev:1997sz, Fleischer:1998dw, Melnikov:2000qh, Chetyrkin:1999qi}
for the $c$-quark the asymptotic behavior of the on-shell-$\overline{\rm{MS}}$ relation reveals itself in the rather low orders of PT, namely in the second or third order (depending on the normalization point). Therefore, in the modern high-precision studies it is more preferable to use the concept of its running mass with the value extracted  e.g. in \cite{Chetyrkin:2009fv, Chetyrkin:2010ic, Dehnadi:2011gc, Kiyo:2015ufa, Alekhin:2017kpj, Mateu:2017hlz}.

In the case of the $b$-quark the first traces of the asymptotic structure of the ratio $M_b/\overline{m}_b(\overline{m}^2_b)$ is observed at the $\mathcal{O}(a^4_s)$ level \cite{Marquard:2015qpa, Marquard:2016dcn}. However, for an unambiguous response to the question about the number of order of PT starting from which the asymptotic behavior will manifest itself it is necessary to know the value of the correction of the fifth order. But already from the available data, it follows that unlike the running mass at the four-level the pole mass of the bottom quark should be used with care. The values of  $\overline{m}_b(\overline{m}^2_b)$ were obtained  at the $\rm{N^3LO}$ level as the final results of the QCD analysis of the properties of $\Upsilon$ system within the static potential studies (see e.g.  \cite{Penin:2014zaa, Ayala:2014yxa, Ayala:2016sdn, Kiyo:2015ufa, Mateu:2017hlz}), the QCD sum rules (see e.g. 
\cite{Chetyrkin:2009fv, Chetyrkin:2010ic, Dehnadi:2015fra}) and of the production cross-section of the $b\overline{b}$-quarks  in the $e^+e^-$ collisions \cite{Beneke:2014pta}. Also worth mentioning the recent results of the lattice QCD determinations \cite{Bazavov:2018omf}. These lattice results  are stimulating a more careful study of the existing uncertainties in the four-loop on-shell-$\overline{\rm{MS}}$ mass relation for the $b$-quark.

For the $t$-quark the situation is even more intriguing. The definite results of the experimental  analysis of Tevatron and LHC data are expressed through a Monte-Carlo $t$-quark mass, which may be related (though with process-dependent uncertainties) to its pole mass (for the detailed consideration see e.g. \cite{Butenschoen:2016lpz, Corcella:2019tgt, Hoang:2020iah}). The average PDG(20) value of this important quantity, obtained from the recent LHC measurements and the updated Tevatron analysis, is $M_t=172.76 \pm  0.30 ~{\rm GeV}$ \cite{Zyla:2020zbs}. For comparison, recently the LHC value of the pole mass of the $t$-quark with the thorough estimates of the various types of uncertainties was obtained in \cite{Aad:2019mkw} and reads
 $M_t=171.1 \pm  0.4(stat)$ $\pm 0.9(syst)^{+0.7}_{-0.3} (theor) ~{\rm GeV}$. Note that in the process of getting the top-quark mass values the question about inaccuracies of  different Monte-Carlo programs used for analyzing Tevatron and LHC data has become more vivid. This problem is still under careful examinations (see \cite{Baskakov:2017jhb, Nason:2017cxd} and the reviews of \cite{Corcella:2019tgt, Hoang:2020iah}). The arising inaccuracies should be compared with other theoretical errors, which enter into the determinations of both running  and pole top-quark masses \cite{Alekhin:2016jjz, Alekhin:2017kpj, Catani:2020tko}. Moreover, the uncertainties, contributed by at least the first not yet  computed high-order correction to the relation between the pole and running masses, are also of interest \cite{Nason:2017cxd, Beneke:2016cbu}. The study of these effects will be continued in this paper using several approaches for estimating high-order QCD corrections to the on-shell-$\overline{\rm{MS}}$ mass relation.
 
Our  main aim is to analyze the asymptotic structure of the perturbative series for the ratio $M_q/\overline{m}_q(\overline{m}^2_q)$ at the $\mathcal{O}(a^6_s)$ level. To get a feeling for what may be  
the values of the five- and six-loop corrections to 
this ratio, we estimate them using three distinct techniques. After this, we restore the general $n_l$-dependence of these estimates (the previous definite results on this topic 
are presented in brief in \cite{Kataev:2018mob, Kataev:2018fvx, PhD thesis}) and demonstrate its sign-alternating character in $n_l$.  

The outline of our studies is as follows. In Sec.~\ref{sec2} we present the current known four-loop corrections to the on-shell-$\overline{\rm{MS}}$ mass relation for the particular case of the $SU(3)$ color gauge group. Here we especially emphasize the appearance of the contributions proportional to powers of $\pi^2$-terms to
the analytical expressions for coefficients of the ratio (\ref{t^M}) starting to manifest itself from the two-loop level and originating from calculation of $Z^{\rm{OS}}_m$ in the Minkowskian on-shell subtraction scheme (the first emergence of a $\pi^4$-term in $Z^{\overline{\rm{MS}}}_m$ occurs at the $\mathcal{O}(a^4_s)$ level only).

In  Sec.~\ref{sec3} we use the K\"allen-Lehmann type dispersion relation for the ``effective'' spectral function, defined in the Euclidean domain for energies, to model the on-shell $\pi^2$-terms contributing to $Z^{\rm{OS}}_m$  by the analytical continuation $\pi^2$-effects arising upon the transition from the Euclidean to Minkow- skian region.

In Sec.~\ref{sec4} we apply the approach proposed in \cite{Kataev:1995vh} and extended in \cite{Chetyrkin:1997wm} to estimate five- and six-loop corrections $t^M_5$ and $t^M_6$ with partial incorporation of the $\pi^{2n}$-contributions being mentioned above. This approximate  procedure is based on the effective-charges (ECH) method \cite{Grunberg:1982fw}
and on the concept of scheme-invariants \cite{Stevenson:1981vj}. 

Sec.~\ref{sec7} is devoted to the study of the consequences following from the results of \cite{Ball:1995ni}, where the exact numerical values of the  contributions leading in powers of $n_l$ to the coefficients $t^M_k$  were computed from consideration of the diagrams containing an insert of a chain of quark loops into the single gluon line, renormalized the massive quark propagator. Note that within the Naive-Nonabelianization (NNA) procedure utilized by us, this leads to the sign-alternating $n_l$-structure of the five- and six-loop PT corrections. 

Sec.~\ref{sec6} is dedicated to the investigation of the $\mathcal{O}(a^5_s)$ and $\mathcal{O}(a^6_s)$-estimates found with help of the asymptotic renorma- lon-based formula for coefficients of the on-shell-$\overline{\rm{MS}}$ relation, which was previously studied in \cite{Beneke:1994rs, Beneke:2016cbu, Beneke:1998ui, Pineda:2001zq, Hoang:2017suc, Ayala:2019hkn}. Herewith, we consider two variants for fixation of the normalization factor included in this factorial formula (for details see \cite{Beneke:2016cbu} and \cite{Hoang:2017suc}). We demonstrate that using both these ways one can obtain the sign-alternating structure of the five- and six-loop coefficients in (\ref{t^M}). This fact is in full agreement with the outcomes following from the application of the ECH-motivated method and the large-$\beta_0$ analysis. Here we especially emphasize that in contrast to the results of our previous works on this topic \cite{Kataev:2018mob, Kataev:2018fvx, PhD thesis} the sign-alternating structure of the renormalon-based estimates is observed upon attraction of more detailed information on the normalization factor of the renormalon asymptotic formula.

In Sec.~\ref{Discussion} we briefly summarize all the main our results presented in the previous sections and consider the numerical impact of the estimated $\mathcal{O}(a^5_s)$ and $\mathcal{O}(a^6_s)$ terms on the behavior of the on-shell-$\overline{\rm{MS}}$ relation for real heavy quarks.
We show that the application of all methods employed by us leads to the results which are consistent with each other in order of magnitude (on average with a factor two).

For clarity in \ref{A} of this paper we set out 
the key points of the LSM, define the way of finding the LSM-solutions for the terms $t^M_{4, 0}$ and $t^M_{4, 1}$ and 
their uncertainties. Note here that as follows from the studies of \cite{Kataev:2015gvt, Kataev:2018sjv} these solutions of the overdetermined system of algebraic equations are stable under a change not only in the number of $n_l$-equations being considered, but also in the number of unknowns involving in this system.

In order to consider the possible differences in the structure of the perturbative series in QCD and QED in \ref{B} we compare the behavior of the PT series for the relation between the pole and running masses of the heavy quarks in QCD with the corresponding one for the charged leptons in QED at the four-loop level.

\section{The on-shell-$\overline{\rm{MS}}$ heavy quark mass relation: available analytical perturbative QCD results}
\label{sec2}

Consider first the relation (\ref{t^M}) between 
the pole and running heavy quark masses normalized at the scale  $\mu^2=\overline{m}^2_q$. It is known that the heavy quark pole mass $M_q$ is defined in the on-shell scheme as a pole of the renormalized heavy quark propagator in the Minkowski region. In turn, the scale evolution of the $\rm{\overline{MS}}$-scheme heavy quark running mass is first defined in the Euclidean domain since the calculations of the corresponding master integrals for $Z^{\overline{\rm{MS}}}_m$ are also performed in the Euclidean region:
\begin{eqnarray}
\label{ms-5}
\frac{\overline{m}_q(Q^2)}{\overline{m}_q(\mu^2)}&=&{\rm{exp}}\left(\int\limits_{\alpha_s(\mu^2)}^{\alpha_s(Q^2)}dx\;\frac{\gamma_m(x)}{\beta(x)}\right)~.
\end{eqnarray}

This relation may be transformed to the Minkowski region by replacement $Q^2\rightarrow s$. After this, it is possible to fix the Minkowskian scale $\mu^2=\overline{m}_q^2$ and to define $\overline{m}_q(\overline{m}_q^2)$. The dependence of the 
QCD expansion parameter  
$a_s(\mu^2)$ and of the running  quark mass $\overline{m}_q(\mu^2)$ on the renormalization scale $\mu^2$ is 
determined by the following RG equations:
\begin{eqnarray}
\label{RG-eq1}
&&\mu^2\frac{\partial a_s}{\partial \mu^2}=\beta(a_s)=-\sum\limits_{n=0}^{\infty} \beta_n a^{n+2}_s~, \\ 
\label{RG-eq2}
&&\mu^2\frac{\partial \ln(\overline{m}_q)}{\partial \mu^2}=\gamma_m(a_s)=-\sum\limits_{n=0}^{\infty}\gamma_n a^{n+1}_s~,
\end{eqnarray}
where $\beta(a_s)$ and $\gamma_m(a_s)$ are the QCD $\beta$-function and the anomalous mass dimension. 
In our further consideration  we use their   
$\rm{\overline{MS}}$-like  scheme  expressions.  The one- and two-loop coefficients $\beta_0$ and $\beta_1$  of the QCD $\beta$-function  were  computed analytically in  \cite{Gross:1973id, Politzer:1973fx} and \cite{Jones:1974mm, Caswell:1974gg, Egorian:1978zx} respectively. The symbolical expressions of the  scheme-dependent three- and four-loop  coefficients $\beta_2$ and $\beta_3$ are known from  calculations performed in \cite{Tarasov:1980au,  Larin:1993tp} and \cite{vanRitbergen:1997va, Czakon:2004bu} correspondingly. The  coefficient $\beta_4$ was obtained in analytical form in the $SU(3)$-group \cite{Baikov:2016tgj} and confirmed in \cite{Herzog:2017ohr, Luthe:2017ttg} by computing this term in the general $SU(N_c)$ gauge group. Note that in the process of these calculations the Euclidean contribution, proportional to the $\zeta_4=\pi^4/90$ Riemann function, is appearing for the first time. 

For our purposes it is convenient to present these coefficients $\beta_n$ in  terms of the number of massless flavors $n_l=n_f-1$. 
In the case of $SU(3)$ color gauge group  their numerical expressions  have the following form:
\begin{subequations}
\begin{eqnarray}
\label{b0-1}
\beta_0&=&-0.166667n_l+2.58333~, \\
\label{bb}
\beta_1&=&-0.791667n_l+5.58333~, \\
\label{b2}
\beta_2&=&+0.094039n^2_l-4.18084n_l+18.0454~, \\
\label{b3}
\beta_3&=&+0.005857n^3_l+1.5999n^2_l-23.951n_l+88.684~, \\
\label{b4}
\beta_4&=& -0.0017993n^4_l-0.233054n^3_l+16.46765n^2_l \\ \nonumber 
&-&148.1715n_l+359.687~.
\end{eqnarray}
\end{subequations}

The first scheme-independent coefficient $\gamma_0$  of the QCD  anomalous mass dimension function of Eq.(\ref{RG-eq2})
was presented in \cite{Tarrach:1980up}. Its  two-, three- and four-loop expressions 
were analytically  computed in \cite{Tarrach:1980up, Nachtmann:1981zg}, \cite{Tarasov:1982gk, Larinmass}, \cite{Vermaseren:1997fq, Chetyrkin:1997dh} correspondingly. The coefficient $\gamma_4$ of the  fifth order was 
evaluated in case of the $SU(3)$ color gauge group in  \cite{Baikov:2014qja}. This analytical result had been confirmed later on in \cite{Luthe:2016xec} upon more general calculations performed in the $SU(N_c)$-group.  It should be stressed that the Euclidean contributions $\zeta_4$ being proportional to $\pi^4$ are arising in the QCD expression for $\gamma_m$ beginning from the four-loop level (see \cite{Vermaseren:1997fq, Chetyrkin:1997dh}), whereas the functions $\zeta_6$ proportional to $\pi^6$ are starting to reveal themselves at the five-loop level\footnote{For the explanation of the ``postponed'' 
manifestation of the even contributions $\zeta_{2n}$ in the analytical expressions of the QCD RG-functions of Eqs.(\ref{RG-eq1}-\ref{RG-eq2}) see \cite{Baikov:2018wgs}.}.

The numerical values of these coefficients are:
\begin{subequations}
\begin{eqnarray}
\label{gammi}
\gamma_0&=&1~, ~~~~~~
\gamma_1=-0.138889n_l+4.06944~, \\
\gamma_2&=&-0.027006n_l^2-2.33813n_l+17.2045~, \\
\gamma_3&=&+0.00579n^3_l+0.29354n^2_l-18.5378n_l+80.117 \\ \label{gamma4}
\gamma_4&=&-0.0000854n^4_l+0.107977n^3_l+7.80682n^2_l\\ \nonumber
&-&128.3970n_l+423.611~.
\end{eqnarray}
\end{subequations}

Note that all renormalized quantities, which enter into ratio (\ref{t^M}), are self-consistently defined in the Minkowski region of energies.
In particular case of the $SU(3)$ group the  analytical contributions to the first four coefficients of Eq.(\ref{t^M}), expanded in powers of $n_l$ (see Eq.(\ref{zim})), follow from the calculations of \cite{Tarrach:1980up, Gray:1990yh, Melnikov:2000qh, Lee:2013sx} and  read 
\begin{subequations}
\begin{eqnarray}
\label{10} 
t^M_{1, 0}&=&\frac{4}{3}, \\ \label{21}
t^M_{2, 1}&=&-\frac{71}{144}-\frac{\pi^2}{18}, \\ \label{20}
t^M_{2, 0}&=&\frac{307}{32}-\frac{\zeta_3}{6}+\frac{\pi^2}{3}+\frac{\pi^2\ln 2}{9}, \\
\label{32} 
t^M_{3, 2}&=&\frac{2353}{23328}+\frac{7}{54}\zeta_3+\frac{13}{324}\pi^2, \\ \label{31}
t^M_{3, 1}&=&-\frac{231847}{23328}-\frac{241}{72}\zeta_3 
+\frac{\ln^4 2}{81}+\frac{8}{27}{\rm{Li}}_4\bigg(\frac{1}{2}\bigg) \\ \nonumber
&+&\frac{61}{1944}\pi^4 
+\pi^2\bigg[-\frac{991}{648}-\frac{11}{81}\ln 2+\frac{2}{81}\ln^2 2\bigg], \\  \label{30}
t^M_{3, 0}&=&\frac{8481925}{93312}+\frac{58}{27}\zeta_3+\frac{1975}{216}\zeta_5-\frac{55}{162}\ln^4 2 \\ \nonumber
&-&\frac{220}{27}{\rm{Li}}_4\bigg(\frac{1}{2}\bigg)-\frac{695}{7776}\pi^4+\pi^2\bigg[\frac{652841}{38880}-\frac{1439}{432}\zeta_3 \\ \nonumber
&-&\frac{575}{162}\ln 2-\frac{22}{81}\ln^2 2\bigg], \\  \label{43}
t^M_{4, 3}&=&-\frac{42979}{1119744}-\frac{317}{2592}\zeta_3-\frac{71}{25920}\pi^4-\frac{89}{3888}\pi^2, \\
\label{42}
t^M_{4, 2}&=&\frac{30575329}{4478976}+\frac{40979}{5184}\zeta_3-\frac{241}{216}\zeta_5-\frac{11}{486}\ln^4 2 \\ \nonumber
&+&\frac{\ln^5 2}{405}-\frac{44}{81}{\rm{Li}}_4\bigg(\frac{1}{2}\bigg)-\frac{8}{27}{\rm{Li}}_5\bigg(\frac{1}{2}\bigg) \\ \nonumber
&+&\pi^4\bigg[\frac{32293}{466560}+\frac{31}{9720}\ln 2\bigg]+\pi^2\bigg[\frac{6979}{3456}+\frac{5}{48}\zeta_3 \\ \nonumber
&+&\frac{103}{972}\ln 2-\frac{11}{243}\ln^2 2+\frac{2}{243}\ln^3 2\bigg],
\end{eqnarray}
\end{subequations}
where ${\rm{Li}}_n(x)=\sum\limits_{k=1}^{\infty}x^kk^{-n}$ is the polylogarithmic function.

As the result the two- and three-loop coefficients of the ratio $M_q/\overline{m}_q(\overline{m}_q^2)$ have the following numerical form:
\begin{subequations}
\begin{eqnarray}
\label{t2nm}
t^M_2&=&-1.0414n_l+13.443~, \\
\label{t3nnm}
t^M_3&=&+0.6527n_l^2-26.655n_l+190.60~.
\end{eqnarray}

The numerical $n_l$-dependent expression for the $\mathcal{O}(a^4_s)$ term $t^M_4$ is known at present 
with high enough accuracy. We combine here the results of the analytical (\ref{43}-\ref{42}) \cite{Lee:2013sx} and semi-analytical computations \cite{Marquard:2016dcn} with the LSM-solutions \cite{Kataev:2018sjv} for the constant $t^M_{4, 0}$
and linearly dependent on $n_l$ term $t^M_{4, 1}$ with their LSM-uncertainties. This leads to  the following expression (see \ref{A}):
\begin{eqnarray}
\label{t4M}
t^M_4=&-&0.6781n_l^3+43.396n_l^2 \\ \nonumber 
&-&(745.72\pm 0.036)n_l+(3567.61\pm 1.62)~.
\end{eqnarray}
\end{subequations}

Note that for our purposes to study the \textit{asymptotic} structure of the on-shell-$\overline{\rm{MS}}$ mass relation the uncertainties included in (\ref{t4M}) are not important and we can neglect them. 

Unlike the coefficients of the QCD $\beta$-function and  the anomalous mass dimension the results (\ref{t2nm}-\ref{t4M}) clearly de- monstrate the sign-alternating pattern in $n_l$. It is interesting to note that this computational fact is consistent with the  theoretical renormalon-inspired large $\beta_0$-expansion  \cite{Ball:1995ni, Beneke:1994qe}. 

We now return to the discussion concerning the analytical structure of certain contributions to the formulas (\ref{21}-\ref{42}). It is worth emphasizing that the second, third and fourth coefficients $t^M_2$, $t^M_3$, $t^M_4$   
contain the $\pi^2$-terms typical to the Minkowskian on-shell subtraction scheme, while the additional $\pi^4$-contributions are emergering  in the results of three-loop calculations and beyond. We expect the appearance of the $\pi^6$-terms in the structure of analytical expressions for the yet unknown coefficients $t^M_{4, 0}$ and $t^M_{4, 1}$\footnote{This expectation is already supported by the recent QED analytical results of $t^M_{4, 0}$ in \cite{Laporta:2020fog}.}. Comparing the analytical  structure of the $\mathcal{O}(a^4_s)$ perturbative QCD corrections to the ratio $M_q/\overline{m}_q(\overline{m}_q^2)$ and to the  QCD  anomalous mass dimension   $\gamma_m(a_s)$, dictated by the pattern of the quark mass renormalization constant $Z_m^{\overline{\rm{MS}}}$, one can conclude that only the $\pi^4$-contributions, entering into the $t^M_{4, 2}$ and $t^M_{4, 1}$, may contain the admixture of the typical Euclidean $\zeta_4$-terms, 
first appearing in the four-loop contributions to $\gamma_m(a_s)$, which are proportional to $n_l^2$ and $n_l$  \cite{Vermaseren:1997fq, Chetyrkin:1997dh}. Other contributions to the coefficients $t^M_2-t^M_4$, proportional to powers of $\pi^2$, arise from computations of high-order corrections to the  renormalization constant $Z^{\rm{OS}}_m$ defined in Eq.(\ref{defOS}) in the Minkowskian on-shell subtraction scheme.
In next section we try to build an analogy between 
these typical on-shell scheme $\pi^2$-contribu- tions and the ``kinematic'' effects proportional to powers of $\pi^2$ in the perturbative QCD expressions for  
the Minkowskian physical quantities, initially evaluated in the $\overline{\rm{MS}}$-scheme in the Euclidean domain. The substantial role of these effects has been demonstrated in the number of works on the subject (see e.g. \cite{Radyushkin:1982kg, Gorishnii:1983cu, Pivovarov:1991bi, LeDiberder:1992jjr, Altarelli:1994vz, Broadhurst:2000yc, Bakulev:2010gm, Nesterenko:2017wpb}).

\section{Is it possible to link the on-shell $\pi^2$-contributions and ``kinematic'' $\pi^2$-effects?}
\label{sec3}

To understand whether it is possible to draw the analogy between contributions proportional to powers of $\pi^2$ to the ratio $M_q/\overline{m}_q(\overline{m}_q^2)$, which are defined in the Minkowskian region and  the ``kinematic'' $\pi^2$-terms, arising in the PT coefficients of the Minkowskian RG controllable physical quantities in the $\overline{\rm{MS}}$-scheme and associated with the analytical continuation effects from the Euclidean to Minkowskian domain, we will follow the path treaded in \cite{Chetyrkin:1997wm} and used later on in \cite{Kataev:2010zh}. 
For this goal, we consider the K\"allen-Lehmann
type dispersion representation\footnote{
For instance, the similar dispersion relation links the Euclidean Adler function $D(Q^2)$ for a process of $e^+e^-$ annihilation into hadrons with the  $R(s)$-ratio, characterizing the total cross section of this process in the Minkowskian region of energies.}, which allows to  simulate the appearance of these ``kinematic'' terms:
\begin{equation}
\label{dispersion}
F(Q^2)=Q^2\int\limits_{0}^{\infty} ds \frac{T(s)}{(s+Q^2)^2}.
\end{equation}

Here the model spectral function $T(s)$ is determined in the Minkowski region\footnote{The quantity $T(s)$ may be expressed through combinations containing an imaginary part of the self-energy insertions to the renormalized quark propagator considered in \cite{Tarrach:1980up}.} as:
\begin{equation}
\label{T(s)}
T(s)=\overline{m}_q(s)\bigg(1+\sum\limits_{k=1}^{\infty}t_k a^k_s(s)\bigg).
\end{equation} 

In this perturbative expression $\overline{m}_q(s)$ is the $\rm{\overline{MS}}$-scheme running mass of heavy quark, normalized at the scale $\mu^2=s$ in the time-like region  and  $t_k$ are the dimensionless coefficients of this spectral function\footnote{At an arbitrary normalization point the coefficients $t_k$ contain the RG logarithms of a type $\ln (\mu^2/s)$.}. One of the basic ideas of the work \cite{Chetyrkin:1997wm}
consists in a fact that at $s=\overline{m}^2_q$ the coefficients $t_k$ in Eq.(\ref{T(s)}) are assumed to be equal to the corresponding on-shell scheme coefficients $t^M_k$ of the heavy quark mass relation (\ref{t^M}), i.e. at this point $T(\overline{m}^2_q)=M_q$.

Substituting the expression (\ref{T(s)}) into relation (\ref{dispersion}) one can arrive to the perturbative
 representation for the Euclidean function $F(Q^2)$
\begin{equation}
\label{F}
F(Q^2)=\overline{m}_q(Q^2)\bigg(1+\sum\limits_{n=1}^{\infty}f^E_n a^n_s(Q^2)~\bigg)
\end{equation}
with coefficients $f^E_k$ related to $t^M_k$ by the following way 
\begin{equation}
\label{tn-Deltan}
t^{M}_k = f^E_k-\Delta_k,
\end{equation}
where contributions $\Delta_k$ are the ``kinematic'' terms, which reflect the analytic continuation effects.

Note that from the point of view of the first principles of the theory of dispersion representations the model equation (\ref{dispersion}) is not completely substantiated. Indeed, within PT it should contain the subtraction constant, which is related to the theoretical ambiguities in the low-energy region, discussed in \cite{Broadhurst:2000yc, Pivovarov:2001xj} upon a study of the dispersion representations of the Green's functions for the scalar quark and gluon currents.
In this regard, it would be more consistent to consider the model subtracted dispersion relation written down for the function $F(Q^2)-F(0)$. However, below we will show that the perturbative estimates for coefficients of the ratio $M_q/\overline{m}_q(\overline{m}^2_q)$, obtained with help of the expression (\ref{dispersion}), yield the quite reasonable predictions of the asymptotic behavior of this ratio and agree with applications of the renormalon-motivated calculus (with a factor of order 2). 

Keeping in mind the aforesaid discussions, Eqs.(\ref{dispersion}-\ref{T(s)}), remark on equality of coefficients $t_k$ and $t^M_k$ at  $s=\overline{m}^2_q$
and taking into account the inverse integral representation for the function $T(s)$\footnote{By analogy with the dispersion relation between the $e^+e^-$ annihilation $R(s)$-ratio and the Adler $D(Q^2)$-function, the integration contour on the plane of complex variable $z$ lies in the region of analyticity of the integrand (here function $F(Q^2)$ is the analog of the Adler function).} 
\begin{equation}
\label{2pii}
T(s)=\frac{1}{2\pi i}\int\limits_{-s-i\varepsilon}^{-s+i\varepsilon} F(z)\frac{dz}{z},
\end{equation}
one can obtain the following approximate representation for the pole and $\rm{\overline{MS}}$-scheme masses of heavy quarks \cite{Chetyrkin:1997wm}:
\begin{equation}
\label{Massdispersion}
 M_q \approx \frac{1}{2\pi i}\int\limits_{-\overline{m}_q(\overline{m}^2_q)-i\epsilon}^{-\overline{m}_q(\overline{m}^2_q)+i\epsilon}
ds^{\prime}\displaystyle\int\limits_0^{\infty}\frac{\overline{m}_q(s)(1+\sum_{k=1}^{\infty}t^M_k a^k_s(s))}{(s+s^{\prime})^2}ds
\end{equation}

Using now Eqs.(\ref{dispersion}-\ref{F}) we can fix the
 explicit form of the ``kinematic'' contributions $\Delta_k$
  in (\ref{tn-Deltan}) up to the sixth order of PT. Far enough from the regions of manifestation of the heavy quark threshold effects the differential system of RG-equations (\ref{RG-eq1}-\ref{RG-eq2})  in the time-like region can be rewritten in the following integral form in the $\mathcal{O}(a^6_s)$ approximation:
\begin{equation}
\left\{
\begin{aligned}
&\ln\frac{\mu^2}{s}=\hspace{-0.25cm}\int\limits_{a_s(\mu^2)}^{a_s(s)}\hspace{-0.25cm} \frac{dx}{\beta_0 x^2+\beta_1 x^3+\beta_2 x^4+\beta_3 x^5+\beta_4 x^6+\beta_5 x^7}~, \\ \nonumber
&\ln\frac{\overline{m}_q(s)}{\overline{m}_q(\mu^2)}
=\hspace{-0.25cm}\int\limits_{a_s(\mu^2)}^{a_s(s)}  \hspace{-0.2cm}\frac{(\gamma_0 +\gamma_1 x+\gamma_2 x^2+\gamma_3 x^3+\gamma_4 x^4+\gamma_5 x^5)dx}{\beta_0 x+\beta_1 x^2+\beta_2 x^3+\beta_3 x^4+\beta_4 x^5+\beta_5 x^6}
\end{aligned}
\right.
\end{equation}

Substituting solutions of this system into function $T(s)$ in Eqs.(\ref{dispersion}-\ref{T(s)}) we get the following integrals which are equal to:
\begin{eqnarray}
\label{integration}
&&Q^2\int\limits_{0}^{\infty}ds\frac{\{1; l; l^2; l^3; l^4; l^5; l^6\}}{(s+Q^2)^2}=
\bigg\{1;~ \mathfrak{L};~ \mathfrak{L}^2+\frac{\pi^2}{3};  \\ \nonumber 
&&\mathfrak{L}^3+\pi^2\mathfrak{L}; ~\mathfrak{L}^4+2\pi^2\mathfrak{L}^2+\frac{7\pi^4}{15}; ~\mathfrak{L}^5+\frac{10}{3}\pi^2\mathfrak{L}^3+\frac{7}{3}\pi^4\mathfrak{L}; \\ \nonumber 
&&\mathfrak{L}^6+5\pi^2\mathfrak{L}^4+7\pi^4\mathfrak{L}^2+\frac{31}{21}\pi^6\bigg\}~,
\end{eqnarray}
where $l=\ln (\mu^2/s)$ and $\mathfrak{L}=\ln (\mu^2/Q^2)$.
Fixing further $\mu^2=Q^2$  we find the explicit expressions for the terms $\Delta_k$.  In the recurrent form they read:
\begin{subequations}
\begin{eqnarray}
\label{Eu-Minkowski}
&&\Delta_1=0~~,~~~\Delta_2=\frac{\pi^2}{6}\gamma_0(\beta_0+\gamma_0)~, \\
\label{D3}
&&\Delta_3=\frac{\pi^2}{3}\bigg[f^E_1(\beta_0+\gamma_0)\bigg(\beta_0+\frac{1}{2}\gamma_0\bigg) +\frac{1}{2}\beta_1\gamma_0 \\ \nonumber
&&+\gamma_1\beta_0+\gamma_1\gamma_0\bigg]~, \\
\label{D4} 
&&\Delta_4=\frac{\pi^2}{3}\bigg[(f^E_2-\Delta_2)\bigg(3\beta^2_0+\frac{5}{2}\beta_0\gamma_0+\frac{1}{2}\gamma_0^2\bigg) \\ \nonumber
&&+f^E_1\bigg(\frac{3}{2}\beta_1\gamma_0+\frac{5}{2}\beta_1\beta_0+2\gamma_1\beta_0+\gamma_1\gamma_0\bigg) \\ \nonumber
&&+\frac{1}{2}\beta_2\gamma_0+\gamma_1\beta_1+\frac{1}{2}\gamma^2_1+\frac{3}{2}\gamma_2\beta_0+\gamma_2\gamma_0\bigg] \\ \nonumber
&&+\frac{7\pi^4}{60}\gamma_0(\beta_0+\gamma_0)\bigg(\beta_0+\frac{1}{2}\gamma_0\bigg)\bigg(\beta_0+\frac{1}{3}\gamma_0\bigg)~, \\ 
\label{D5}
&&\Delta_5=\frac{\pi^2}{3}\bigg[(f^E_3-\Delta_3)\bigg(6\beta^2_0+\frac{7}{2}\beta_0\gamma_0+
\frac{1}{2}\gamma^2_0\bigg) \\ \nonumber
&&+(f^E_2-\Delta_2)\bigg(7\beta_1\beta_0+3\gamma_1\beta_0 
+\frac{5}{2}\beta_1\gamma_0+\gamma_1\gamma_0\bigg) \\ \nonumber 
&&+f^E_1\bigg(\frac{3}{2}\beta^2_1+\frac{1}{2}\gamma^2_1+3\beta_2\beta_0+\frac{5}{2}\gamma_2\beta_0 
+2\beta_1\gamma_1 \\ \nonumber
&&+\frac{3}{2}\beta_2\gamma_0+\gamma_2\gamma_0\bigg)
+\frac{1}{2}\beta_3\gamma_0+\beta_2\gamma_1+\frac{3}{2}\gamma_2\beta_1+2\gamma_3\beta_0 \\ \nonumber
&&+\gamma_1\gamma_2+\gamma_0\gamma_3\bigg]+\frac{7\pi^4}{15}\bigg[f^E_1\bigg(\beta^4_0+\frac{25}{12}\beta^3_0\gamma_0+\frac{35}{24}\beta^2_0\gamma^2_0
\\ \nonumber
&&+\frac{5}{12}\beta_0\gamma^3_0+\frac{1}{24}\gamma^4_0\bigg)+\gamma_1\beta^3_0+\frac{13}{12}\gamma_0\beta_1\beta^2_0+\frac{13}{12}\gamma^2_0\beta_0\beta_1 \\ \nonumber
&&+\frac{11}{6}\gamma_0\gamma_1\beta^2_0+\gamma^2_0\beta_0\gamma_1+\frac{1}{4}\beta_1\gamma^3_0+\frac{1}{6}\gamma_1\gamma^3_0\bigg]~, \\
\label{D6}
&&\Delta_6=\frac{\pi^2}{3}\bigg[(f^E_4-\Delta_4)\bigg(10\beta^2_0+\frac{9}{2}\beta_0\gamma_0+\frac{1}{2}\gamma^2_0\bigg) 
\\ \nonumber 
&&+(f^E_3-\Delta_3)\bigg(\frac{27}{2}\beta_0\beta_1+4\beta_0\gamma_1+\frac{7}{2}\beta_1\gamma_0+\gamma_0\gamma_1\bigg) \\ \nonumber
&&+(f^E_2-\Delta_2)\bigg(8\beta_0\beta_2+\frac{7}{2}\beta_0\gamma_2+3\beta_1\gamma_1 +\frac{5}{2}\beta_2\gamma_0+4\beta^2_1 \\
\nonumber
&&+\frac{1}{2}\gamma^2_1+\gamma_0\gamma_2\bigg)+f^E_1\bigg(\frac{7}{2}\beta_0\beta_3+\frac{7}{2}\beta_1\beta_2+3\beta_0\gamma_3 \\ \nonumber
&&+\frac{5}{2}\beta_1\gamma_2+2\beta_2\gamma_1+\frac{3}{2}\beta_3\gamma_0+\gamma_0\gamma_3+\gamma_1\gamma_2\bigg)+\frac{1}{2}\gamma^2_2+\frac{3}{2}\beta_2\gamma_2 \\ \nonumber
&&+\frac{5}{2}\beta_0\gamma_4+2\beta_1\gamma_3+\beta_3\gamma_1+\frac{1}{2}\beta_4\gamma_0+\gamma_0\gamma_4+\gamma_1\gamma_3\bigg] \\ \nonumber
&&+\frac{7\pi^4}{15}\bigg[(f^E_2-\Delta_2)\bigg(5\beta^4_0+\frac{77}{12}\beta^3_0\gamma_0 +\frac{71}{24}\beta^2_0\gamma^2_0\\ \nonumber
&&+\frac{7}{12}\beta_0\gamma^3_0+\frac{1}{24}\gamma^4_0\bigg)
+f^E_1\bigg(\frac{77}{12}\beta^3_0\beta_1+\frac{5}{12}\beta_1\gamma^3_0 
\end{eqnarray}
\begin{eqnarray}
 \nonumber
&&+4\beta^3_0\gamma_1+\frac{1}{6}\gamma^3_0\gamma_1+\frac{10}{3}\beta_0\beta_1\gamma^2_0+\frac{25}{3}\beta^2_0\beta_1\gamma_0+\frac{3}{2}\beta_0\gamma^2_0\gamma_1 \\ \nonumber
&&+\frac{13}{3}\beta^2_0\gamma_0\gamma_1\bigg)+\frac{1}{4}\beta_2\gamma^3_0+\frac{5}{2}\beta^3_0\gamma_2+\frac{1}{6}\gamma^3_0\gamma_2+\frac{3}{2}\beta^2_0\gamma^2_1 \\ \nonumber
&&+\frac{5}{8}\beta^2_1\gamma^2_0+\frac{1}{4}\gamma^2_0\gamma^2_1+\frac{35}{24}\beta_0\beta^2_1\gamma_0+\frac{5}{4}\beta_0\beta_2\gamma^2_0+\frac{47}{12}\beta^2_0\beta_1\gamma_1 \\ \nonumber
&&+\frac{3}{2}\beta^2_0\beta_2\gamma_0+\frac{5}{4}\beta_0\gamma_0\gamma^2_1+\frac{5}{4}\beta_0\gamma^2_0\gamma_2+\beta_1\gamma^2_0\gamma_1+\frac{37}{12}\beta^2_0\gamma_0\gamma_2 \\ \nonumber
&&+\frac{25}{6}\beta_0\beta_1\gamma_0\gamma_1\bigg]+\frac{31\pi^6}{126}\gamma_0(\beta_0+\gamma_0)\bigg(\beta_0+\frac{1}{2}\gamma_0\bigg)\times \\ \nonumber
&&\times\bigg(\beta_0+\frac{1}{3}\gamma_0\bigg)\bigg(\beta_0+\frac{1}{4}\gamma_0\bigg)\bigg(\beta_0+\frac{1}{5}\gamma_0\bigg)~.
\end{eqnarray}
\end{subequations}

The terms $\Delta_1-\Delta_4$ agree with the ones, obtained previously in \cite{Chetyrkin:1997wm}. The expressions  for $\Delta_5$ and  $\Delta_6$ are new.
One can see that the six-loop contribution $\Delta_6$ does not contain yet unknown coefficients $\beta_5$ and $\gamma_5$. They are included only in terms depending linearly on $\ln(\mu^2/s)$, which
due to Eq.(\ref{integration})
vanish automatically in the Euclidean renormalization point $\mu^2=Q^2$. 

Taking now into account the relation (\ref{tn-Deltan}) and 
numerical expressions for the coefficients of 
$\beta(a_s)$, $\gamma_m(a_s)$ and $t^M_k$, given in Eqs.(\ref{b0-1}-\ref{gamma4}), (\ref{t2nm}-\ref{t4M}),
we arrive to the following  numerical 
$n_l$-dependent results for the terms $\Delta_k$:
\begin{subequations}
\begin{eqnarray}
\label{C2}
\Delta_2&=&-0.274156n_l+5.89434, \\ 
\Delta_3&=&+0.198002n^2_l-10.04477n_l+105.6221, \\ \label{C4}
\Delta_4&=&-0.315898n^3_l+20.67673n^2_l-403.9489n_l \\ \nonumber
&+& 2272.002, \\ 
\label{C5}
\Delta_5&=&+0.427523n^4_l-37.745285n^3_l+1137.17794n^2_l \\ \nonumber
&-&13767.2725n_l+56304.639, \\ 
\label{C6}
\Delta_6&=&-0.818446n^5_l+85.37937n^4_l 
-3345.0818 n^3_l \\ \nonumber
&+& 61128.1667 n^2_l 
- 518511.694 n_l+1633115.62,
\end{eqnarray}
\end{subequations}
where in the expression for $\Delta_6$-contribution we have neglected the relatively small mean square errors following from computations of the coefficient $t^M_4$ \cite{Marquard:2016dcn}. 

Worth emphasizing that despite the non-regular sign polynomial structure of the coefficients of the QCD 
RG functions $\beta(a_s)$ and $\gamma_m(a_s)$ 
(\ref{b0-1}-\ref{gamma4}), the analogous expressions for contributions $\Delta_k$ respect the alternation of signs in powers of $n_l$ that is typical to the two, three and four-loop coefficients $t^M_k$.  
 
Their numerical values for the specific number of massless flavors are presented in Table~\ref{Table4}:
\vspace{-0.5cm}
\begin{table}[!h]
\begin{center}
\small
{\def\arraystretch{1.2}\tabcolsep=0.1pt
\begin{tabular}{|c|c|c|c|c|c|}
\hline 
$~~~n_l~~~$ &  ~~~~~$\Delta_2$~~~~~ & ~~~~~~~$\Delta_3$~~~~~~ & ~~~~~~~~~$\Delta_4$~~~~~~~~ & ~~~~~~~~~$\Delta_5$~~~~~~~~~ & ~~~~~~~~~$\Delta_6$~~~~~~~~~ \\
\hline 
3 &    5.072 & 77.270 & 1237.717 & 24252.930 & 544133.68 \\
\hline
4 &    4.798 & 68.611 & 966.817 & 17124.144  &  344053.30\\
\hline
5 &    4.524 & 60.348 & 729.689 & 11446.766  &  201430.55 \\
\hline
\end{tabular}}
\end{center}
\caption{The numerical values of the $\Delta_k$-contributions.}
\label{Table4}
\end{table}

\vspace{-0.5cm}
The outcomes of Table~\ref{Table4} demonstrate  the significant \\ growth of terms $\Delta_k$ with increasing of an order $k$ of PT. This effect is determined by two factors. 
The first of them is related to the factorial rise of the coefficients $t^M_k$ included in the definition of the contributions $\Delta_k$ (\ref{D3}-\ref{D6}) (see Sec.~\ref{sec6} of this paper, where the renormalon-based asymptotic formula for $t^M_k$ is discussed). The second one is partially associated with the considerable factorial growth of the constant terms appearing in r.h.s of Eq.(\ref{integration}) upon the integration of the RG-logarithms with various degrees.
 Indeed, as was shown in \cite{Bjorken:1989xw} the dimensionless analog of integral (\ref{integration})
with arbitrary degree $n$ has the 
closed form
\begin{eqnarray}
\label{closed}
&&\int\limits_0^{\infty}dx\frac{\ln^n x}{(x+1)^2}= \left\{
\begin{aligned}
&2(1-2^{1-n})\zeta_nn!~, ~~~ n ~~\text{even}~, \\
&0~, ~~~~~~~~~~~~~~~~~~~~~~~~~~~ n ~~\text{odd}~,
\end{aligned}
\right.
\end{eqnarray}
where the variable $x=s/Q^2$ and terms $\zeta_2=\pi^2/6$, $\zeta_4=\pi^4/90$, $\zeta_6=\pi^6/945$ 
may be explicitly restored in r.h.s of Eq.(\ref{integration}). Since $1<\zeta_{2p}\leq \zeta_2<2$ for any $p\in\mathbb{N}$, then at even $n$ the integral (\ref{closed}) is factorially growing. As a result, the constant terms in  r.h.s of Eq.(\ref{integration}), which enter in the contributions $\Delta_k$, are  
factorially increasing with order of PT as well.
 Moreover, matching Eqs.(\ref{Eu-Minkowski}-\ref{D6}) with (\ref{closed}) we conclude that the contribution to the even-order term $\Delta_{2p}$ leading in powers of $\pi^2$  behaves itself by the following way for any $p\in\mathbb{N}$:
\begin{eqnarray}
\Delta_{2p}^{\text{max}\;\pi}&=&2(1-2^{1-2p})\zeta_{2p}(2p-1)!\gamma_0\prod_{j=1}^{2p-1}\bigg(\beta_0+\frac{\gamma_0}{j}\bigg) \\ \nonumber
&=&2(1-2^{1-2p})\zeta_{2p}\gamma_0\beta_0^{2p-1}\frac{\Gamma(2p+\gamma_0/\beta_0)}{\Gamma(1+\gamma_0/\beta_0)}~.
\end{eqnarray}

Thus, we conclude that the overall Minkowskian ``kinematic'' $\pi^2$-effects are indeed not small. Moreover, the values of $\Delta_2$, $\Delta_3$, $\Delta_4$ are comparable with the corresponding coefficients $t^M_k(n_l)$ (see Eqs.(\ref{Mcharm}-\ref{Mtop}) in \ref{B}). In this regard and in view of our assumption that these fast growing ``kinematic'' effects may model the $\pi^2$-contributions to the high-order coefficients of $M_q/\overline{m}_q(\overline{m}_q^2)$-ratio, typical to the on-shell renormalization scheme, we note that it is really worth to treat these special terms with care.

\section{The effective charges-inspired estimates} 
\label{sec4}

Let us  first study the application of a variant of an RG-inspired approach for estimating high-order perturbative corrections to a physical quantities being formulated and developed in \cite{Kataev:1995vh}. This approach is based on the  effective-charges (ECH) method  \cite{Grunberg:1982fw}.  In the work \cite{Chetyrkin:1997wm} it was first  adapted to the quantity $F(Q^2)/\overline{m}_q(Q^2)$ 
defined in the Euclidean region.
Since here we consider the case of $n_l$ massless flavors, running inside the fermion loop insertions of a self-energy operator renormalizing the massive heavy quark propagator, then the coefficients $f^E_n$ in Eq.(\ref{F}) do not depend on masses. In this approximation the perturbative expression for $F(Q^2)/\overline{m}_q(Q^2)$ is also  independent on mass parameter. Therefore one can use directly the methods described in 
\cite{Kataev:1995vh, Grunberg:1982fw}.
The corresponding ECH coupling constant $a^{eff}_s(Q^2)$ may be defined as:
\begin{eqnarray}
\label{effective} 
\frac{F(Q^2)}{\overline{m}_q(Q^2)}&=&1+f^E_1 a^{eff}_s(Q^2)~, \\ 
\label{phik}
 a^{eff}_s(Q^2)&=&a_s(Q^2)+\sum\limits_{k=2}^{\infty}\phi_k a^k_s(Q^2)~, 
\end{eqnarray} 
where $\phi_k=f^E_k/f^E_1$. The coefficients of the  ECH $\beta$-function for  $a^{eff}_s(Q^2)$ are expressed through scheme-independent combinations \cite{Stevenson:1981vj} of the higher order PT contributions $\phi_k$ (\ref{phik}) and $\beta_k$ of the $\rm{\overline{MS}}$-scheme $\beta$-function. At the four-loop level these combinations have already  
been applied for determination of the ECH $\beta$-function of the static potential in the QCD \cite{Kataev:2015yha}.
Here we present the explicit expressions for six  coefficients of the corresponding  
ECH $\beta$-function,  which is governing the $Q^2$-behavior of $a^{eff}_s(Q^2)$: 
\begin{subequations}
\begin{eqnarray}
\beta^{eff}_0&=&\beta_0~, ~~~~~~\beta^{eff}_1=\beta_1~, \\
\label{betaef0-2}
\beta^{eff}_2&=&\beta_2-\phi_2\beta_1+(\phi_3-\phi^2_2)\beta_0~, \\
\beta^{eff}_3&=&\beta_3-2\phi_2\beta_2+\phi^2_2\beta_1+(2\phi_4-6\phi_2\phi_3+4\phi^3_2)\beta_0, \\
\label{betaef4}
\beta^{eff}_4&=&\beta_4-3\phi_2\beta_3+(4\phi^2_2-\phi_3)\beta_2+(\phi_4-2\phi_2\phi_3)\beta_1 \\ \nonumber
&+&(3\phi_5-12\phi_2\phi_4-5\phi^2_3+28\phi^2_2\phi_3-14\phi^4_2)\beta_0~, \\ 
\label{betaef5}
\beta^{eff}_5&=&\beta_5-4\phi_2\beta_4+(8\phi^2_2-2\phi_3)\beta_3 \\ \nonumber
&+&(4\phi_2\phi_3-8\phi^3_2)\beta_2  +(2\phi_5-8\phi_2\phi_4-3\phi^2_3 \\ \nonumber
&+& 16\phi^2_2\phi_3-6\phi^4_2)\beta_1
+(4\phi_6-20\phi_2\phi_5-16\phi_3\phi_4 \\ \nonumber
&+&48\phi_2\phi^2_3-120\phi^3_2\phi_3+56\phi^2_2\phi_4+48\phi^5_2)\beta_0~.
\end{eqnarray}
\end{subequations}

Our further analysis is based on the  theoretical studies described in \cite{Chetyrkin:1997wm, Kataev:1995vh, Kataev:2010zh}. Their essence was as follows: if one put $\beta^{eff}_2\approx\beta_2$, then from Eq.(\ref{betaef0-2}) one can get that $f^E_3\approx(f^E_2)^2/f^E_1+f^E_2\beta_1/\beta_0$, where $f^E_2=t^M_2+\Delta_2$. After this, using the additional contribution $\Delta_3$, responsible for the transition from the Euclidean to Minkowski region (\ref{Eu-Minkowski}), one can fix the approximate  value of $t^M_3$-term (we denote it as $t^{M,\;ECH}_3$). Similarly, supposing  $\beta^{eff}_3\approx\beta_3$ one can estimate the value of the four-loop contributions $f^E_4$ and  $t^{M, \; ECH}_4$ afterwards. Estimates of this type were made in \cite{Chetyrkin:1997wm, Kataev:2010zh} to fix the numerical value of the term $t^M_4$ for the cases of the charm, bottom and top-quarks which was still unknown at that time. 

An admissibility of the approximation $\beta^{eff}_k \approx \beta_k$ in the asymptotic regime is somewhat supported by the effect of partial cancellation of the renormalon contributions in the coefficients of $\beta^{eff}_k$ (\ref{betaef0-2}-\ref{betaef5}) and their absence in the QCD $\beta$-function.

Note also that in principle one may apply 
the ECH-based  estimating procedure  in the time-like  region directly. In this case one should change 
the Euclidean functions in (\ref{effective}-\ref{phik}) to their  Minkowski counterparts $T(s)/\overline{m}_q(s)$,  $a_s^{eff}(s)$ and $a_s(s)$ with coefficients $\phi^{M}_k=t^M_k/t^M_1$ instead of $\phi_k$. Then after applying the main ansatz of the ECH-based procedure in the time-like region, namely 
$\beta^{eff,\;M}_k \approx \beta_k$, we will get the estimates of $t^M_k$-terms directly without 
additional evaluation of the ``kinematic'' $\Delta_k$-corrections ($t^{M,\; ECH\; direct}_k$ stands for these estimates).
Nevertheless, these terms will 
include the $\pi^2$-contributions typical to the on-shell scheme. 
Indeed, in the estimates $t^{M,\; ECH\; direct}_k$ of $k$-th order these $\pi^2$ effects are contained in 
the known analytical Minkowskian coefficients of $(k-1)$-th order and lower. This approach will be considered in more details below.

The estimates $t^{M,\; ECH}_3$, $t^{M, \;ECH\; direct}_3$, $t^{M, \;ECH}_4$,  $t^{M, \;ECH\; direct}_4$, obtained by these ways, are compared to their exact expressions $t^{M,\;exact}_3$
and $t^{M, \;exact}_4$ (see Eqs.(\ref{t3nnm}-\ref{t4M})) in Table~\ref{Table1}.
\vspace{-0.3cm}
\begin{table}[!h]
\begin{center}
{\def\arraystretch{1.2}\tabcolsep=0.1pt
\begin{tabular}{|c|c|c|c|}
\hline 
$~~n_l~~$ & ~~~~$t^{M,\;exact}_3$~~~~  & ~~~~$t^{M, \;ECH}_3$~~~~ & ~~~~$t^{M, \;ECH\; direct}_3$~~~~  \\
\hline 
3 & $\;$ 116.494 $\;$ &  124.097  & 95.757   \\
\hline
4 & 94.418 & 97.728 & 76.257   \\
\hline
5 & 73.637 & 73.615 & 58.528   \\
\hline
6 & 54.161 & 51.775 & 42.615   \\
\hline
7 & 35.991 & 32.235 & 28.583   \\
\hline
8 &  19.126 & 15.034 & 16.535  \\
\hline
\hline
$~~n_l~~$ & ~~$t^{M, \;exact}_4$~~ & ~~$t^{M, \;ECH}_4$~~ & ~~$t^{M, \; ECH\; direct}_4$~~ \\
\hline
3 & $\;$ $1702.70\pm 1.62$  $\;$ & $\;$ 1281.09 $\;$ & 1438.76 \\
\hline
4 & $1235.66\pm 1.63$ & 986.13  & 1045.51 \\
\hline
5 & $839.14\pm 1.63$ & 719.38 & 710.02    \\
\hline
6 & $509.07\pm 1.63$ & 483.02 & 430.94    \\
\hline
7 & $241.37\pm 1.64$ & 279.37 & 207.02    \\
\hline
8 & $31.99\pm 1.65$   & 110.71 & 37.19    \\
\hline
\end{tabular}}
\end{center}
\caption{The exact values and estimates of coefficients $t^M_3$ and $t^M_4$.}
\label{Table1}
\end{table}

\vspace{-0.3cm}
One can see from data of  Table~\ref{Table1} that  both variants of the ECH-motivated method give quite good approximations for the three- and four-loop coefficients of the ratio $M_q/\overline{m}_q(\overline{m}_q^2)$\footnote{Note that the values of the corrections $t^{M,\; ECH}_3$ and $t^{M,\; ECH\; direct}_3$, presented in Table~\ref{Table1}, are slightly different from the analogous ones, obtained by means of the same ECH-motivated method in \cite{Kataev:2010zh}.  The discrepancy between them lies in a slip made in \cite{Kataev:2010zh}. However, this fact did not affect the final results of the four-loop terms $t^{M, \; ECH}_4$ and $t^{M, \; ECH\; direct}_4$.} (apart from the non-physical case of $n_l=8$ for the ECH approach, where the estimation differs from the genuine value by a factor over 3).  Indeed, both these implementations predict not only the correct signs for the coefficients of the $\mathcal{O}(a^3_s)$ and $\mathcal{O}(a^4_s)$ orders but also yield the estimates whose values are rather close to the expressions having been calculated exactly.

Let us now probe the $n_l$-dependence of the estimated coefficient $t^{M}_3$ with three unknowns $t^M_{3,0}$, $t^M_{3,1}$, $t^M_{3,2}$ with help of three physical data points $3\leq n_l\leq 5$. Solving the corresponding system of equations we gain the following expansions:
\begin{subequations}
\begin{eqnarray}
\label{t3esten}
t^{M,\; ECH}_3&\approx &+1.128n^2_l-34.265n_l+216.74~, \\ \label{t3estdm}
t^{M,\; ECH\; direct}_3&\approx &+0.885n^2_l-25.69n_l+164.87~.
\end{eqnarray} 
\end{subequations}

The approximate results (\ref{t3esten}-\ref{t3estdm}) are in good agreement with the genuine one (\ref{t3nnm}). Furthermore, they keep the sign-alternating structure in powers of $n_l$ as well.

Similarly, extracting the flavor dependence of the estimates $t^{M, \;ECH}_4$ and $t^{M, \; ECH\; direct}_4$ 
from the systems of four equations,
formed by numerical values at  $n_l=3, 4, 5, 6$, we arrive to the following decompositions:
\begin{subequations} 
\begin{gather}
\label{t4estE}
t^{M, \; ECH}_4\approx +0.36n^3_l+9.75n^2_l-376.62n_l+2313.43~, \\ 
\label{t4estM}
t^{M,\; ECH\; direct}_4\approx -0.224n^3_l+31.56n^2_l-605.9n_l \\ \nonumber 
~~~~~~~~~~~~~~~~~~~~~~~~+2978.44~.
\end{gather} 
\end{subequations}

\begin{figure}[h!]
\includegraphics[width=0.48\textwidth]{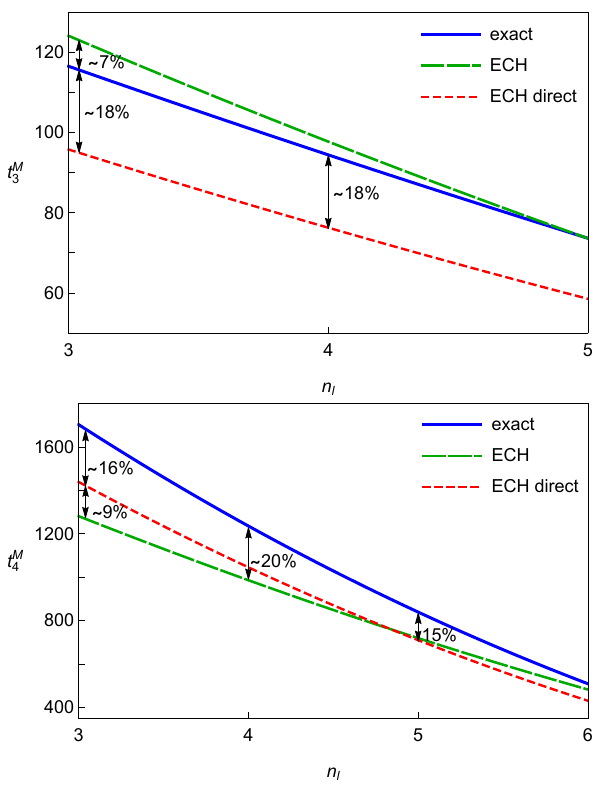}
\caption{\label{gr1} The flavor dependence of the terms $t^M_3$ and $t^M_4$. The exact results (blue line) and the approximate ones, obtained by both variants of the ECH-inspired method (with and without the explicit supplementation of the analytic continuation effects (dashed green and red lines correspondingly)), are presented.
} 
\end{figure}

The sign-alternating structure of the $n_l$-expanded expression (\ref{t4estM}), gotten in the Minkowskian region directly, is consistent with  result (\ref{t4M}) of the explicit diagram-by-diag- ram calculations. However, the ECH approach, applied in the Euclidean region, leads to the
different sign of the leading cubic term in (\ref{t4estE}). To clarify this observation we note that its value almost coincides modulo with $n_l^3$-contribution to $\Delta_4$ in Eq.(\ref{C4}). This means that the cubic coefficient in $f^E_4=t^M_4+\Delta_4$ is close to zero. For a more detailed study of this fact we present Figure~\ref{gr1}, where the 
obtained expansions (\ref{t3esten}-\ref{t3estdm}),
(\ref{t4estE}-\ref{t4estM}) are visually compared to the exact results (\ref{t3nnm}-\ref{t4M}). 

It is seen from Figure~\ref{gr1} that the relative uncertainties of the ECH-direct approach are stable to the changes of $n_l$. However, this is not true for the Euclidean ECH method applied for estimation of the term $t^M_4$. Indeed, in this case the relative error varies in a wide range from $25\%$ at $n_l=3$ to $5\%$ at $n_l=6$. Moreover, at $n_l=8$ we observe a mismatch with the exact result by a factor over 3 (see Table~\ref{Table1}). Therefore, we conclude that at the relative errors of about $25\%$, one should not trust the almost zero  estimated value
of the cubic $n_l$-dependent term in $f^E_4$ coefficient. Thus, the errors of this order are quite satisfactory while getting the estimates of the term $t^M_4$ (or $f^E_4$) at the fixed number of massless quarks but they turn out to be unsatisfactory for the study of more subtle effects of its flavor dependence. Therefore, we infer that the mismatch of the sign of $n^3_l$-term in Eq.(\ref{t4estE}) to the true one is a rather accidental fact related to the instability of the uncertainties being discussed above. In view of this we do not consider the positive sign of this cubic coefficient as a violation of the indication of the sign-alternating structure of the $n_l$-expanded ECH-based estimates of $t^M_4$-term following from (\ref{t4estM}).

The acceptable agreement of the ECH-estimates of the coefficients $t^M_3$ and $t^M_4$ with the results of explicit calculations at the fixed number of $n_l$ allows us to regard both variants of the ECH-inspired method as satisfactory
estimating  procedures. Therefore, we will apply these two realizations to evaluate the unknown contributions of the fifth and sixth orders of PT to the on-shell-$\overline{\rm{MS}}$ heavy quark mass relation as well.

Our further studies of the ECH method, initially applied to the Euclidean physical quantities, will contain the following steps:
\begin{enumerate}
\item 
At the first stage, using the explicit expressions for the known terms $t^M_1-t^M_4$ (\ref{10}), (\ref{t2nm}-\ref{t4M}) and adding to them $\pi^2$-effects of the analytical continuation (\ref{C2}-\ref{C4}), we find the Euclidean contributions $f^E_1-f^E_4$.
\item 
Secondly, fixing $\beta^{eff}_4\approx\beta_4$ (this is our main guess) we get from Eq.(\ref{betaef4}) the approximate form of the $f^E_5$-term for specific values of $n_l$.
\item
Then, accordingly to (\ref{tn-Deltan}) we subtract from the obtained term $f^E_5$ the contribution of analytic continuation $\Delta_5$ (\ref{C5}) and get $\mathcal{O}(a^5_s)$ coefficient $t^M_5$.
\item
Applying this procedure in the next order of PT, i.e 
assuming $\beta^{eff}_5\approx\beta_5$ in (\ref{betaef5}) and using the numerical expression for $f^E_5$, obtained at the previous stage\footnote{In view of this the uncertainties in the definition of the six-loop corrections to the mass conversion formula will certainly be greater than for the five-loop ones.}, we primarily estimate the value of $f^E_6$-contribution and then, taking into account (\ref{tn-Deltan}) and (\ref{C6}),  evaluate the value of $t^M_6$-correction to the ratio $M_q/\overline{m}_q(\overline{m}_q^2)$.
\end{enumerate}

In stages 2 and 4, we get the following five- and six-loop  coefficients $f^E_5$ and $f^E_6$ of the Euclidean quantity (\ref{F}):
\begin{subequations}
\begin{eqnarray}
\label{f^E_5}
f^E_5&\approx &\frac{1}{3\beta_0}\bigg[3f^E_2\beta_3+\bigg(f^E_3-4\frac{(f^E_2)^2}{f^E_1}\bigg)\beta_2 \\ \nonumber
&+&\bigg(2\frac{f^E_2f^E_3}{f^E_1}-f^E_4\bigg)\beta_1\bigg]+4\frac{f^E_2f^E_4}{f^E_1}+\frac{5}{3}\frac{(f^E_3)^2}{f^E_1} \\ \nonumber
&-&\frac{28}{3}f^E_3\bigg(\frac{f^E_2}{f^E_1}\bigg)^2+\frac{14}{3}\frac{(f^E_2)^4}{(f^E_1)^3}~, \\
\label{f^E_6}
f^E_6&\approx &\frac{1}{4\beta_0}\bigg[4f^E_2\beta_4+\bigg(2f^E_3-8\frac{(f^E_2)^2}{f^E_1}\bigg)\beta_3 \\ \nonumber
&+&\bigg(8\frac{(f^E_2)^3}{(f^E_1)^2}-4\frac{f^E_2f^E_3}{f^E_1}\bigg)\beta_2+\bigg(6\frac{(f^E_2)^4}{(f^E_1)^3}+3\frac{(f^E_3)^2}{f^E_1} \\ \nonumber
&+&8\frac{f^E_2f^E_4}{f^E_1}-16f^E_3\bigg(\frac{f^E_2}{f^E_1}\bigg)^2-2f^E_5\bigg)\beta_1\bigg] \\ \nonumber
&+&5\frac{f^E_2f^E_5}{f^E_1}+4\frac{f^E_3f^E_4}{f^E_1}+30f^E_3\bigg(\frac{f^E_2}{f^E_1}\bigg)^3 \\ \nonumber
&-&12f^E_2\bigg(\frac{f^E_3}{f^E_1}\bigg)^2
-12\frac{(f^E_2)^5}{(f^E_1)^4}-14f^E_4\bigg(\frac{f^E_2}{f^E_1}\bigg)^2~.
\end{eqnarray}
\end{subequations}

Utilizing these expressions and Eqs.(\ref{C5}-\ref{C6})
for $\Delta_5$, $\Delta_6$ that model the ``kinematic'' $\pi^2$-terms, we  estimate the ECH-values of the coefficients $t^M_5$, $t^M_6$
at the fixed number of massless flavors given in Table~\ref{Table3}. There we also present the estimates of the same coefficients, obtained by us from the application of the large-$\beta_0$ expansion to the
renormalon-chain contributions \cite{Ball:1995ni} to  coefficients of the on-shell-$\overline{\rm{MS}}$  mass relation (see 4- and 5-th columns of Table~\ref{Table3}) 
and with the help of the infrared renormalon (IRR) asymptotic formula \cite{Beneke:1998ui, Pineda:2001zq, Beneke:2016cbu},  used recently in \cite{Mateu:2017hlz, Hoang:2017suc, Komijani:2017vep} (see 6-th column). The details of these our analyzes will be discussed below.

\begin{table}[!h]
\begin{center}
{\def\arraystretch{1.2}\tabcolsep=0.1pt
\begin{tabular}{|c|c|c|c|c|c|}
\hline 
$~n_l~$ & ~~~~$t^{M, \; ECH}_5$~~~~ & ~~~$t^{M, \; ECH\;direct}_5$~~~  & 
~~~~$t^{M, \; FL}_5$~~~~  & ~~~$t^{M, \; FL,\; M\rightarrow\overline{m}}_5$~~~~ & ~~~~$t^{M, \; r-n}_5$~~~~ \\
\hline 
3 & 28435 & 26871  & 29864 & 20432 & 33859  \\
\hline
4 & 17255 & 17499  & 21951 & 14924 & 22602  \\
\hline
5 & 9122  & 10427  & 15725 & 10757 & 13942  \\
\hline
6 & 3490  & 5320   & 10929 & 7693  & 7543   \\
\hline
7 & -127  & 1871   & 7323 &  5515  & 3108  \\
\hline
8 & -2153 & -196   & 4693 &  4027  & 321  \\
\hline \hline
$~n_l~ $ & ~~~~$t^{M, \; ECH}_6$~~~~ & ~~~$t^{M, \; ECH\;direct}_6$~~~  & 
~~~~$t^{M, \; FL}_6$~~~~  & ~~~$t^{M, \; FL,\; M\rightarrow\overline{m}}_6$~~~ & ~~~~$t^{M, \; r-n}_6$~~~~\\
\hline 
3 & 476522 & 437146 & 679654 & 522713 & 825382 \\
\hline
4 & 238025 & 255692 & 462561 & 353810 & 507235 \\
\hline
5 & 90739  & 133960 & 304866 & 233282 & 285136 \\
\hline
6 & 8412   & 57920  & 193449 & 149601 & 138664 \\
\hline
7 & -29701 & 15798  & 117284 & 93225  & 50340  \\
\hline
8 & -39432 & -2184  & 67253  & 56410  & 4431 \\
\hline
\end{tabular}}
\end{center}
\caption{The estimates of the coefficients $t^M_5$ and $t^M_6$, obtained within two variants of the ECH-approach ({\it{ECH}} and {\it{ECH~direct}}), 
 the large $\beta_0$-expansion ({\it{FL}} and {\it{FL}}, $M\rightarrow\overline{m}$) and the asymptotic IRR-based formula ({\it{r-n}}) with the normalization factor $N_m$ taken from the direct analysis and consistent with the renormalon sum rule approach (for details see \ref{sec61} and \ref{sec62}). The fourth and fifth columns correspond to various 
choices of the initial scales in the results of \cite{Ball:1995ni} used by us.} 
\label{Table3}
\end{table}

The data from Table~\ref{Table3} demonstrates that at the physical values of $n_l$ the obtained estimates agree with each other at the level of factor two. However, the theoretical uncertainties increase drastically starting from   the non-physical sector $n_l\geq 6$. Indeed, on the contrary to the results following from the application of the NNA procedure to the outcomes of \cite{Ball:1995ni} (see columns 4 and 5) and from the renormalon asymptotic formula \cite{Beneke:1998ui, Beneke:2016cbu, Hoang:2017suc} (see column 6), the ECH-based estimates at $n_l=7,8$ take the  negative values. The similar sign-changing feature also reveals itself in the renormalon studies from $n_l\geq 9$
 (see e.g. the analysis of \cite{Ayala:2014yxa}  and \cite{Beneke:2016cbu}). Moreover, the indication of the sharp growth of the uncertainties in the non-physical sector of $n_l$ also follows directly from the renormalon studies \cite{Hoang:2017suc} (see discussions in Sec.~\ref{sec6} below). Therefore, in this work we restrict ourselves by the consideration of the values of $n_l$ from the range $3\leq n_l\leq 8$. This number of data points is definitely enough to investigate the flavor dependence of the six-loop coefficient $t^M_6$. 
 
Let us study the $n_l$-dependence of the ECH coefficients whose numerical values at the fixed number of $n_l$ are given in Table~\ref{Table3}. As follows from (\ref{zim}) the five-loop contribution $t^M_5$ is the fourth degree polynomial in $n_l$, namely  $t^M_5=t^M_{5, 4}n^4_l+t^M_{5, 3}n^3_l+t^M_{5, 2}n^2_l+t^M_{5, 1}n_l+t^M_{5, 0}$. It contains five unknown terms $t^M_{5, 4}-t^M_{5, 0}$. Therefore, in order to get their numerical values we will use five equations only which follow from the data of Table~\ref{Table3} at $3\leq n_l\leq 7$. Their matrix representation\footnote{The square matrix in l.h.s. of (\ref{system-1}) is the Vandermonde matrix. It possesses the interesting mathematical properties: the elements of its each row are the terms of a geometric progression and its determinant is equal to  $\Delta=\prod\limits_{0\leq i< j\leq k} ((n_l+j)-(n_l+i))=\prod\limits_{0\leq i< j\leq k} (j-i)$. Here the number of massless quarks varies from $n_l$ to $(n_l+k)$, $\; k\in\mathbb{N}$. } read:
\vspace{-0.3cm}
\begin{equation}
\label{system-1}
\begin{pmatrix}
    1 & 3 & 9 & 27 & 81  \\
    1 & 4 & 16 & 64 & 256 \\
    1 & 5 & 25 & 125 & 625 \\
    1 & 6 & 36 & 216 & 1296 \\
    1 & 7 & 49 & 343 & 2401  \\
   \end{pmatrix}
   \begin{pmatrix}
   t^{M, \; ECH}_{5, 0} \\ 
   t^{M, \; ECH}_{5, 1} \\
   t^{M, \; ECH}_{5, 2} \\
   t^{M, \; ECH}_{5, 3} \\
   t^{M, \; ECH}_{5, 4} \\
   \end{pmatrix}
   = \begin{pmatrix} 
   28435 \\
   17255 \\
   9122 \\
   3490  \\
   -127  \\
   \end{pmatrix}
   \end{equation}
   
The numerical solution of (\ref{system-1}) leads to the following expression:
\begin{subequations}
\begin{equation}
\label{t^MECH_5}
t^{M, \; ECH}_5\approx 2.5n^4_l-136n^3_l+2912n^2_l-26976n_l+86620
\end{equation}

In the case of the coefficient of the sixth order of PT the similar consideration at $3\leq n_l\leq 8$ yields:
\begin{eqnarray}
\label{t^MECH_6}
t^{M, \; ECH}_6&\approx & -4.9n^5_l +352n^4_l-9708n^3_l \\ \nonumber
&+&131176n^2_l-855342n_l+2096737~.
\end{eqnarray}

Both expansions (\ref{t^MECH_5}-\ref{t^MECH_6}) have 
the sign-alternating structure in powers of $n_l$, which is supported by results of the large-$\beta_0$ analysis 
\cite{Ball:1995ni}. Thus, the ECH-motivated method, applied initially in the Euclidean domain and supplemented by the analytical continuation $\pi^2$-effects, leads to the $n_l$-dependent structure of the terms $t^M_5$ and $t^M_6$, which is similar to the ones observed for the exactly calculated corrections $t^M_2-t^M_4$ given in (\ref{t2nm}-\ref{t4M}).

Let us consider what will happen with the expressions (\ref{t^MECH_5}-\ref{t^MECH_6}) if one fix in them $t^M_{5,4}\approx 0.9$ and $t^M_{6,5}\approx -1.5$, following from the exact numerical computations of Ref.\cite{Ball:1995ni}. These 
results were obtained there from the consideration of the subset of the specific renormalon-chain diagrams for the heavy quark propagator. Since in this case one less $n_l$-dependent term should be defined, we exclude from the analysis the data points $n_l=7(8)$ upon the estimation of the flavor dependencies of the coefficients $t^{M, \;ECH}_{5(6)}$ correspondingly. This leads to the insignificant changes in all coefficients of the expressions (\ref{t^MECH_5}-\ref{t^MECH_6}) with keeping their sign-alternating character\footnote{The result of this analysis gives $t^{M, \; ECH}_{5, \; fixed \; n^4_l}\approx 0.9n^4_l-107n^3_l+2723n^2_l-26429n_l+86041;$ ~ $t^{M, \; ECH}_{6, \; fixed \; n^5_l}\approx -1.5n^5_l +267n^4_l-8873n^3_l+127178n^2_l-845982n_l+2088184$.}. 

Using numbers shown in the third column of Table~\ref{Table3}, we also obtain the approximate $n_l$-dependence of the
$\mathcal{O}(a^5_s)$ and $\mathcal{O}(a^6_s)$ 
 coefficients of the on-shell-$\rm{\overline{MS}}$ heavy quark mass relation within the ECH-motivated approach, applied directly in the Minkowskian region:
\begin{eqnarray}
\label{M-ECH-direct-5}
t^{M, \; ECH\;direct}_5&\approx &1.2n^4_l-77n^3_l 
+1959n^2_l-20445n_l \\ \nonumber
&+&72557~, \\
\label{M-ECH-direct-6}
t^{M, \; ECH\;direct}_6&\approx &-2.2n^5_l+148n^4_l-4561n^3_l \\ \nonumber
&+&71653n^2_l-538498n_l+1519440~.
\end{eqnarray}
\end{subequations}

Despite the definite numerical discrepancy in the values of the five- and six-loop coefficients $t^{M, \; ECH}_k$ and $t^{M, \; ECH\; direct}_k$, especially in the nonphysical sector of $n_l$ (see Table~\ref{Table3}), both 
realizations of the ECH method predict not only the sign-alternating structure of these corrections in powers of $n_l$ but also lead to the values of the separate $n_l$-dependent terms close in magnitude. Note also that the fixation of the known terms leading in $n_l$ does not substantially affect the values of other coefficients in (\ref{M-ECH-direct-5}-\ref{M-ECH-direct-6})\footnote{\label{12}In this case we should compare them with the  expressions $t^{M, \; ECH\; direct}_{5, \; fixed \; n^4_l}\approx 0.9n^4_l-72n^3_l+1927n^2_l-20354n_l+72461;$ ~ $t^{M, \; ECH\; direct}_{6, \; fixed \; n^5_l}\approx -1.5n^5_l +132n^4_l-4398n^3_l+70870n^2_l-536662n_l+1517760$.}.

In the next sections we will compare these results with the similar ones which follow from the large-$\beta_0$ approximation \cite{Ball:1995ni} and from the asymptotic renormalon formula \cite{Beneke:1994rs, Beneke:1998ui} subsequently improved in \cite{Pineda:2001zq, Beneke:2016cbu, Hoang:2017suc}.

\section{The consequences of the leading renormalon chain calculations}
\label{sec7}

Before the analytical computations \cite{Melnikov:2000qh, Lee:2013sx} of the leading $\mathcal{O}(a^3_s)$ and $\mathcal{O}(a^4_s)$ $n_l$-contributions 
to the coefficients $t^M_3$, $t^M_4$ (see (\ref{32}) and (\ref{43})), these terms were evaluated numerically in \cite{Ball:1995ni}. These results follow from calculations of the leading renormalon-type contributions, generated by a chain of the fermion loop (FL) insertions into the gluon line, renormalizing massive heavy quark propagator. The outcomes of Ref.\cite{Ball:1995ni} contain not only the leading $\mathcal{O}(a^3_s)$ and $\mathcal{O}(a^4_s)$ terms but the analogous ones up to the ninth order as well. Applying the NNA procedure one can estimate the numerical values of the total multiloop contributions to the ratio $M_q/\overline{m}_q(\overline{m}_q^2)$ within the large-$\beta_0$ expansion and get their flavor dependencies. Since the terms leading in $n_l$ do not depend on $\mu^2$, we will consider the five and six-loop estimates in two scale normalizations, namely $\mu^2=\overline{m}^2_q$ and $\mu^2=M^2_q$  with its subsequent transition to the running mass.

Using the results of work \cite{Ball:1995ni} and assuming the normalization at $\mu^2=\overline{m}^2_q$, we  get the following expansions:
\begin{subequations} 
\begin{eqnarray}
\label{t5-F-L}
t^{M, \; FL}_5&\approx &0.9n^4_l-59n^3_l+1469n^2_l-16156n_l+66641, \\
\label{t6-F-L}
t^{M, \; FL}_6&\approx &-1.5n^5_l+125n^4_l-4127n^3_l+68088n^2_l \\ \nonumber
&-&561727n_l+1853698~.
\end{eqnarray}

Next,  presuming that the initial normalization point is fixed on the pole mass and then it is shifted 
to the running one, we find the following analogs of  (\ref{t5-F-L}-\ref{t6-F-L}):
\begin{eqnarray}
\label{t5-F-L-M-m}
t^{M, \; FL,\; M\rightarrow\overline{m}}_5&\approx &0.9n^4_l-56n^3_l+1256n^2_l \\ \nonumber
&-&12383n_l+47721~, \\
\label{t6-F-L-M-m}
t^{M, \; FL, \; M\rightarrow\overline{m}}_6&\approx &-1.5n^5_l+120n^4_l-3779n^3_l \\ \nonumber
&+& 58846n^2_l-460910n_l+1468466~.
\end{eqnarray}
\end{subequations}

These expressions demonstrate that the FL-method supplemented by the NNA procedure gives the strict alternation of signs in the polynomial flavor decomposition of the terms $t^M_5$ and $t^M_6$. This feature is the direct consequence of the application of the large $\beta_0$-expansion. Indeed, in this approximation the $k$-th term $t^M_k$ is proportional to $\beta^{k-1}_0(n_l)$-factor, where the first coefficient of the QCD $\beta$-function is defined in (\ref{b0-1}). Therefore, this approach will always lead to the sign-alternating $n_l$-structure of the estimated corrections in all orders of PT. Moreover, this statement does not depend on the normalization point. Thus, the FL-approach supports the results of the ECH-method presented by us above.

Note that the specific $n_l$-dependent terms in (\ref{t5-F-L-M-m}-\ref{t6-F-L-M-m}) are smaller than the corresponding ones in (\ref{t5-F-L}-\ref{t6-F-L}). Herewith, the latter are closer to those found with help of the ECH-motivated method in (\ref{t^MECH_5}-\ref{t^MECH_6}), (\ref{M-ECH-direct-5}-\ref{M-ECH-direct-6}). The numerical values of the corresponding FL-estimates at the fixed number of $n_l$ are presented in the fourth and fifth column of Table~\ref{Table3}.

\vspace{-0.8cm}
\section{Renormalon-based estimating procedure}
\label{sec6}

Let us now move on to the consideration of another approach for estimation of the high-order corrections to the relation between the pole and $\rm{\overline{MS}}$ running masses of heavy quarks based on the renormalon analysis. It is  known that the ratio $M_q/\overline{m}^2_q(\overline{m}^2_q)$  contains the linear  infrared renormalon (IRR) contributions, which lead to the rather strong  factorial increase of the coefficients in this asymptotic PT series \cite{Bigi:1994em, Beneke:1994sw, Beneke:1998ui}. This fast growth of $t^M_k$-terms is governed by the leading $u=1/2$ IRR pole in the Borel image of the ratio being discussed. The study of the behavior of $t^M_k$-coefficients in the renormalon language  results in the following asymptotic formula  derived in \cite{Beneke:1994rs, Beneke:1998ui, Beneke:2016cbu, Beneke:1998rk}:
\begin{eqnarray}
\label{renormalon-dom}
t^{M, \; r-n}_k &&\xrightarrow{k\rightarrow\infty} \pi N_m (2\beta_0)^{k-1}\frac{\Gamma(k+b)}{\Gamma(1+b)}\bigg(1+\frac{s_1}{k+b-1} \\ \nonumber
&&+\frac{s_2}{(k+b-1)(k+b-2)} \\ \nonumber
&&+\frac{s_3}{(k+b-1)(k+b-2)(k+b-3)}+\mathcal{O}\bigg(\frac{1}{k^4}\bigg)\bigg)~,
\end{eqnarray}
where $\Gamma(x)$ is the Euler Gamma-function, $b=\beta_1/(2\beta^2_0)$ and  the values of the sub-leading coefficients $s_k$, evaluated in  \cite{Beneke:1998ui, Pineda:2001zq, Beneke:2016cbu, Hoang:2017suc}, are presented below. In the finite order of PT the factor $N_m$ depends on $n_l$  and $k$. Note that our normalizations and notations for the coefficients of the QCD $\beta$-function (\ref{b0-1}-\ref{b4}) differ from those used in \cite{Beneke:1994sw, Beneke:1994rs,  Beneke:2016cbu, Pineda:2001zq} upon studying the formula (\ref{renormalon-dom}). Indeed, in these works the analytical expression for the first coefficient of the RG $\beta$-function of the $SU(3)$ QCD is defined as $\beta_0=11/4-n_l/6$, while we are using  $\beta_0=11/4-(n_l+1)/6$ (see (\ref{b0-1})). To coordinate these notations and use directly the asymptotic formula (\ref{renormalon-dom}) we  
need to perform a shift $n_l\rightarrow (n_l-1)$ in Eqs.(\ref{b0-1}-\ref{b4}). In this section we will work in these designations. 

The corresponding expressions for the coefficients $s_k$ read:
\vspace{-0.3cm}
\begin{subequations}
\begin{eqnarray}
s_1&=&\frac{1}{4\beta^4_0}(\beta^2_1-\beta_0\beta_2), 
\end{eqnarray}
\begin{eqnarray}
s_2&=&\frac{1}{32\beta^8_0}(\beta^4_1-2\beta^3_1\beta^2_0-2\beta^2_1\beta_2\beta_0
+4\beta_1\beta_2\beta^3_0 \\ \nonumber
&+&\beta^2_2\beta^2_0-2\beta_3\beta^4_0), \\
s_3&=&\frac{1}{384\beta^{12}_0}(\beta^6_1-6\beta^5_1\beta^2_0+8\beta^4_1\beta^4_0
-3\beta^4_1\beta_2\beta_0 \\ \nonumber
&+& 18\beta^3_1\beta_2\beta^3_0 
-24\beta^2_1\beta_2\beta^5_0 -6\beta^2_1\beta_3\beta^4_0
-12\beta_1\beta^2_2\beta^4_0 \\ \nonumber
&+&3\beta^2_1\beta^2_2\beta^2_0+16\beta_1\beta_3\beta^6_0
-\beta^3_2\beta^3_0+8\beta^2_2\beta^6_0 \\ \nonumber
&+&6\beta_2\beta_3\beta^5_0
-8\beta_4\beta^7_0).
\end{eqnarray}
\end{subequations}

One should also mention that another (recurrent) way of obtaining the formula (\ref{renormalon-dom}) was considered
in \cite{Komijani:2017vep}. It was based on the fact that 
the leading renormalon contribution to the
relation between the pole and running masses of heavy quarks is independent on the $\rm{\overline{MS}}$-scheme mass $\overline{m}_q$ (for details see \cite{Beneke:1994rs} and \cite{Hoang:2008yj, Hoang:2017suc}). On the one hand, this fact allows to use the approximate relation $dM_q/d\overline{m}_q(\overline{m}_q^2)\approx 1$. On the other hand, the RG-based form of this derivative can be obtained from Eq.(\ref{t^M}) with taking into account the running of the coupling constant. Matching the unit to this RG-based expression one can obtain the recurrence relation which results in to the factorial formula (\ref{renormalon-dom}).

The ways to fix values of the normalization factor $N_m$ in a specific finite order of PT are different in various works. 
For instance, in \cite{Ayala:2014yxa, Beneke:2016cbu} they were extracted from juxtaposition the results of the diagram-by-diagram calculations of the coefficients $t^M_k$ with the ones being rewritten in the asymptotic form (\ref{renormalon-dom}). In the works \cite{Campanario:2003ix, Pineda:2001zq, Ayala:2014yxa, Hoang:2017suc, Hoang:2017btd} the approximate analytical expressions for $N_m$ were obtained out of the analysis of the behavior of the Borel image of the ratio (\ref{t^M}). In this paper we will utilize both these ways for fixing of $N_m$-factor. 

The extraction of its accurate values for a certain number of $n_l$ is extremely important for the investigation of the subtle effects of the $n_l$-structure of the coefficients $t^M_k$. As we will show further the change in the accuracy of the used values of $N_m$ from two accounted decimal digits to three ones affects considerably this flavor structure. This understanding is supported by the existence of the renormalon sum rule expressions for the normalization factor $N_m$ \cite{Pineda:2001zq, Hoang:2017btd}\footnote{We are grateful to V. Mateu and A. Pineda for pointing us these results lying beyond the $\beta_0$-approximation.}.

\subsection{The direct analysis}
\label{sec61}

In the first approach, which we will call the direct one, the numerical $\mathcal{O}(a^4_s)$ values of $N_m$ were found in \cite{Beneke:2016cbu} from comparison the results of the explicit four-loop computations \cite{Marquard:2016dcn}
with the numerical expressions that follow from the large $k$ renormalon-based expectations (\ref{renormalon-dom}). 
The results of this analysis are presented in the second 
row of Table~\ref{Table2} and are labeled as \textit{direct}.
Note that they are consistent with the ones obtained in \cite{Ayala:2014yxa}. In the third line of this Table the numerical values of $N_m$, defined within the renormalon sum rule approach \cite{Hoang:2017suc} (see discussions below), are given.
\begin{table}[!h]
\begin{center}
{\def\arraystretch{1.2}\tabcolsep=0.1pt
\begin{tabular}{|c|c|c|c|c|c|c|}
\hline 
$~~~~~n_l~~~~~$ & ~~~~~ 3 ~~~~~  & ~~~~~ 4 ~~~~~ & ~~~~~ 5 ~~~~~ & ~~~~~ 6 ~~~~~ & ~~~~~ 7 ~~~~~ & ~~~~~ 8 ~~~~~  \\
\hline
$~~N_m ~(direct)~~$ &  0.537 & 0.506 & 0.462 & 0.394 & 0.279 & 0.056  \\
\hline
$~~N_m ~(sum~ rule)~~$ &  \begin{tabular}[c]{@{}c@{}} 0.526\\  $\pm$\\ 0.012\end{tabular} & \begin{tabular}[c]{@{}c@{}} 0.492\\  $\pm$\\ 0.016\end{tabular} & \begin{tabular}[c]{@{}c@{}} 0.446\\  $\pm$\\ 0.024\end{tabular} & \begin{tabular}[c]{@{}c@{}} 0.381\\  $\pm$\\ 0.038\end{tabular} & \begin{tabular}[c]{@{}c@{}} 0.271\\  $\pm$\\ 0.063\end{tabular} & \begin{tabular}[c]{@{}c@{}} 0.053\\  $\pm$\\ 0.097\end{tabular}
\\
\hline 
\end{tabular}}
\end{center}
\caption{The values of $N_m$, obtained in the direct and renormalon sum rule approaches in the four-loop approximation at $3\leq n_l\leq 8$.}
\label{Table2}
\end{table}

The comparison of the direct four-loop values of $N_m$, presented in Table~\ref{Table2}, with the three-loop ones,
extracted in a similar way in \cite{Beneke:2016cbu} 
 using the $\mathcal{O}(a^3_s)$ results of calculations \cite{Melnikov:2000qh, Chetyrkin:1999qi},
demonstrates that at least at the fixed physical number of flavors the values of $N_m$-factor are rather stable to the transition from one order of PT to another. Indeed, at $n_l=3, 4, 5$ the relative difference between them does not exceed 15\%, but increases substantially for the non-physical flavors. This observation yields us grounds to expect that at least at $3\leq n_l\leq 5$
the application of the renormalon-based asymptotic formula (\ref{renormalon-dom}) with $N_m$, taken in the four-loop approximation, will lead to the quite acceptable estimates for terms $t^M_5$ and $t^M_6$. However, to investigate the $n_l$-structure of these terms in our further analysis we will also consider the non-physical sector with $6\leq n_l\leq 8$.  The corresponding values of $N_m$-factor are also presented in Table~\ref{Table2}. The numerical estimates of the coefficients $t^M_5$ and $t^M_6$, obtained by the application of the asymptotic formula (\ref{renormalon-dom}),  are provided in the sixth column of 
Table~\ref{Table3}, where they are labeled as $t^{M, \; r-n}_5$ and $t^{M, \; r-n}_6$. 

Using the values of these terms from Table~\ref{Table3}
 we  arrive to  the following  expressions
\begin{subequations}
\begin{eqnarray}
\label{t5-r-n-n}
t^{M, \; r-n}_5&\approx &1.6n^4_l-85n^3_l+2164n^2_l \\ \nonumber
&-&23534n_l+87157~, \\ 
\label{t6-r-n-n}
t^{M, \; r-n}_6&\approx &-9.9n^5_l+372n^4_l-8052n^3_l+115164n^2_l \\ \nonumber
&-& 883651n_l+2629567~,
\end{eqnarray}
\end{subequations}
which keep the sign-alternating $n_l$-structure of the five- and six-loop estimates  we have already established upon application both variants of the ECH-motivated method 
and from the large-$\beta_0$ analysis. The expansions (\ref{t5-r-n-n}-\ref{t6-r-n-n}) are in satisfactory agreement with those given in (\ref{t^MECH_5}-\ref{t^MECH_6}), (\ref{M-ECH-direct-5}-\ref{M-ECH-direct-6}), (\ref{t5-F-L}-\ref{t6-F-L}) and (\ref{t5-F-L-M-m}-\ref{t6-F-L-M-m})\footnote{The incorporation in the numerical analysis the known terms $t^M_{5,4}$ and $t^M_{6,5}$ leads to the following decompositions: $t^{M, \; r-n}_{5, \; fixed \; n^4_l}\approx 0.9n^4_l-72n^3_l+2078n^2_l-23286n_l+86896;$ ~ $t^{M, \; r-n}_{6, \; fixed \; n^5_l}\approx -1.5n^5_l +160n^4_l-5977n^3_l+105216n^2_l-860334n_l+2608231$.}.

However, these results disagree with the sign-non-alter- nating ones, obtained in our previous works on this topic \cite{Kataev:2018mob, Kataev:2018fvx} (see also \cite{PhD thesis}), where the correct signs of the known terms leading in $n_l$ were not even reproduced. The reason for this is the accounting of two decimal digits only in the direct definition of $N_m$-values. Although this approximation is permissible (within the renormalon sum rule error bands shown in Table~\ref{Table2}), it leads to the violation of the changeability of signs in the flavor structures of the $\mathcal{O}(a^5_s)$ and $\mathcal{O}(a^6_s)$ renormalon estimates. Thus, the knowledge of the accurate values of $N_m$ is very important for the study of the fine $n_l$-structure of the coefficients $t^M_k$. Note that at the fixed number of massless flavors the renormalon-based estimates obtained in \cite{Kataev:2018mob, Kataev:2018fvx, PhD thesis} are in good agreement with the ones, presented in this our work.

\subsection{The renormalon sum rule approach}
\label{sec62}

Consider now the values of $N_m$, given in the third line of Table~\ref{Table2} and fixed in work \cite{Hoang:2017suc} by the renormalon sum rule:
\begin{equation}
\label{Nm-sum-rule}
N_m=\frac{2\beta_0}{\pi}\Gamma(1+b)\sum\limits_{k=0}^{\infty}\frac{S_k}{\Gamma(1+b+k)},
\end{equation}
where the expressions for the terms $S_k$ contain the high-order coefficients of the QCD $\beta$-function and the corrections $t^M_k$ of the ratio (\ref{t^M}). This formula of 
\cite{Hoang:2017suc} was obtained from the consideration of the Borel image of the relation between the pole quark mass and its MSR-mass:
\begin{equation}
\label{MSR}
M_q-m^{{\rm{MSR}}}_q(R)=R\sum\limits_{k=1}^{\infty}t^{{\rm{MSR}}}_k(n_l)a^k_s(n_l, R).
\end{equation}

The concept of the low-scale MSR-scheme quark mass $m^{{\rm{MSR}}}_q(R)$ is related to the $\rm{\overline{MS}}$-mass but may be evolved to the renormalization scale $R$ below  mass of the heavy quark being considered (for details see \cite{Hoang:2008yj, Hoang:2017suc}). The MSR-mass coincides with the ${\rm{\overline{MS}}}$ running mass at $R=\overline{m}_q(\overline{m}_q)$. An important feature of (\ref{MSR}) is that unlike the relation between the pole and running quark masses this expression contains the linear term in $R$.  This linear dependence allows to get the Borel transformation function of (\ref{MSR}) in analytical form and the analytical expression (\ref{Nm-sum-rule}) for normalization factor $N_m$ in particular \cite{Hoang:2017suc}. Moreover, as was mentioned in this cited paper, this feature permits to speed up the convergence rate of the series (\ref{Nm-sum-rule}) compared to the analogous one, obtained in  \cite{Pineda:2001zq, Pineda:2017uby} as a residue of the Borel transformation function of the ratio $M_q/\overline{m}_q(\overline{m}^2_q)$ in the leading IRR pole $u=1/2$. Note that the study of the effects of other IR and UV renormalons was considered recently for e.g. in \cite{Ayala:2019hkn}.

The corresponding uncertainties of $N_m$-factor presented in Table~\ref{Table2} were extracted in \cite{Hoang:2017suc} from the scale variation of $R$, which model an effect of the missing higher orders of PT to the ratio (\ref{t^M}). These uncertainties increase with the growth of $n_l$. Indeed, at $n_l=3$ the relative error of $N_m$ makes up 2.3\% and at $n_l=7$ it is about 23.2\%. At $n_l=8$ the absolute error rises sharply and already exceeds the central value of $N_m$. This is one more argument why we do not include in our analysis the non-physical sector with $n_l\geq 8$ upon defining the flavor dependencies of the terms $t^M_5(t^M_6)$.

It is apparent from the data of Table~\ref{Table2} that within the error bands the values of $N_m$, obtained in the renormalon sum rule approach, are consistent with those found by the direct way. Therefore, the expansions (\ref{t5-r-n-n}-\ref{t6-r-n-n}) may be valid for this method as well\footnote{One should note that within the uncertainties demonstrated in Table~\ref{Table2} there exists only the narrow error band for which the solutions of the corresponding systems of equations at $3\leq n_l\leq 7(8)$  will be sign-non-alternating in $n_l$.}. 

At the fixed number  $n_l=3, 4, 5$ the numerical estimates $t^{M, \; r-n}_5$ and $t^{M, \; r-n}_6$, given in Table~\ref{Table3}, are in good agreement with the ones obtained in \cite{Hoang:2017suc}. Moreover, in the case of the $b$-quark 
 our five-loop estimates are consistent with those obtained in \cite{Mateu:2017hlz} from the global fits of the energy of the $Q\bar{Q}$ bound states.

\section{Discussion}
\label{Discussion}

In this section we briefly summarize all main results obtained above and consequences following from them. 
Accordingly to the outcomes presented in the previous sections, there are indications that the flavor structure of the five- and six-loop coefficients in the relation between the pole and running masses of heavy quarks has the sign-alterna- ting character in powers of $n_l$, analogous to the one observed for the two-, three- and four-loop coefficients 
having been calculated exactly. It should be noted in passing
that this behavior agrees with the results of the large-$\beta_0$ expansion.

Comparing the magnitudes of the estimated leading terms in $n_l$, obtained by us in two variants of the ECH-method and with help of the IRR-based formula, we conclude that the results (\ref{M-ECH-direct-5}-\ref{M-ECH-direct-6}) in more extent corresponds to the known coefficients $t^M_{5,4}$ and $t^M_{6,5}$ (see also the related expressions presented in footnote \ref{12}).

\begin{figure}[h!]
\includegraphics[width=0.48\textwidth]{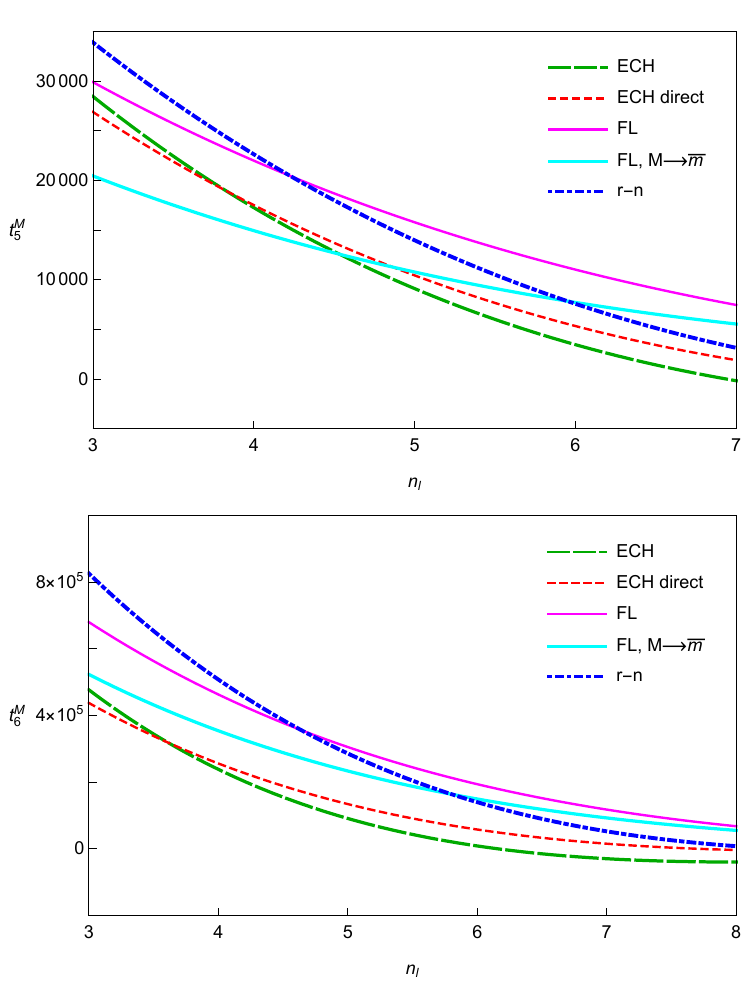}
\caption{\label{gr2} These plots illustrate the flavor 
dependencies of the terms $t^M_5$ and $t^M_6$, obtained within both realizations of the ECH-inspired procedure 
 (with and without the explicit supplementation of analytic continuation effects (the dashed green and red lines correspondingly)), the FL-method, based on the application of the NNA procedure (the solid magenta curve for the direct normalization on the running mass and the cyan line for the initial normalization on the pole mass and the subsequent transition to the running one) and the asymptotic renormalon formula (the dotdashed dark blue line).
} 
\end{figure}

For more clarity, we accompany the estimated flavor dependencies (\ref{t^MECH_5}-\ref{t^MECH_6}), (\ref{M-ECH-direct-5}-\ref{M-ECH-direct-6}), 
(\ref{t5-F-L}-\ref{t6-F-L}), (\ref{t5-F-L-M-m}-\ref{t6-F-L-M-m}), (\ref{t5-r-n-n}-\ref{t6-r-n-n}) of the coefficients $t^M_5$ and $t^M_6$
with the corresponding plots, presented in Figure~\ref{gr2}.

Figure~\ref{gr2} shows that the estimates, obtained in this work with help of all different methods considered by us,  are qualitatively consistent with each other (on average 
with a factor two) and lead to the rather similar flavor structures of the coefficients $t^M_5$ and $t^M_6$. 

Let us now consider an impact of the 
$\mathcal{O}(a^5_s)$ and $\mathcal{O}(a^6_s)$ QCD estimations gotten within all studied approaches on the behavior of the PT series for real heavy quarks in more detail. For numerical studies we will use the central values of the following average PDG(20) numbers \cite{Zyla:2020zbs} for the running masses of $c$ and $b$-quarks, namely $\overline{m}_c(\overline{m}^2_c)=1.27\pm 0.02\; {\rm{GeV}}$,  $\overline{m}_b(\overline{m}^2_b)=4.18^{+ 0.03}_{-0.02} \; \rm{GeV}$. 
 
In accordance with the results of \cite{Ball:2014uwa}
 obtained from the LHC $t\overline{t}$ experimental data and given  in \cite{Alekhin:2016jjz}, for $t$-quark we assume $\overline{m}_t(\overline{m}^2_t)=164.3\pm 0.6 \; \rm{GeV}$  
that does not contradict the values of the running $t$-quark mass presented in PDG(20).

As the initial normalization point we take the average value of the strong coupling constant normalized on the mass of $Z$-boson $\alpha_s(M^2_Z)=0.1179$ at $M_Z=91.1876\; {\rm{GeV}}$ from PDG(20).  Thence from the inverse logarithmic representation of $\alpha_s(M^2_Z)$ we obtain the following value of the scale parameter for the $b$-quark $\Lambda^{(n_l=4)}_{\overline{\rm{MS}}}=207\;\rm{MeV}$, obtained in the four-loop ($\rm{N^3LO}$) approximation.
The numerical results for $\Lambda^{(n_l=3)}_{\overline{\rm{MS}}}$ and $\Lambda^{(n_l=5)}_{\overline{\rm{MS}}}$  are defined using the corresponding $\rm{N^3LO}$ matching transformation conditions, derived  in \cite{Chetyrkin:1997sg, Kniehl:2006bg}\footnote{The inclusion in the numerical analysis of the five-loop 
threshold effects investigated in \cite{Kniehl:2006bg, Chetyrkin:2005ia, Schroder:2005hy}
and of the five-loop contribution to the QCD $\beta$-function \cite{Baikov:2016tgj, Herzog:2017ohr} 
does not affect essentially the numerical values of the pole masses of heavy quarks.} (the corresponding $\rm{NNLO}$ conditions were obtained in \cite{Bernreuther:1981sg, Larin:1994va}  and are naturally taken into account by us),  where the matching scales  are fixed by the values of the $\rm{\overline{MS}}$-scheme masses
presented above. Using the corresponding inverse logarithmic $\rm{N^3LO}$ approximation for $\alpha_s$, we find:
\begin{subequations}
\begin{eqnarray}
\label{Lambdan=3}
\Lambda^{(n_l=3)}_{\overline{\rm{MS}}}&=&289 \;{\rm{MeV}}, \;\;\;\; \alpha_s(\overline{m}^2_c)=0.3929, \\
\label{Lambdan=4}
\Lambda^{(n_l=4)}_{\overline{\rm{MS}}}&=&207\;{\rm{MeV}}, \;\;\;\; \alpha_s(\overline{m}^2_b)=0.2246, \\
\label{Lambdan=5}
\Lambda^{(n_l=5)}_{\overline{\rm{MS}}}&=&88\;{\rm{MeV}}, \;\;\;\;\;\; \alpha_s(\overline{m}^2_t)=0.1083.
\end{eqnarray}
\end{subequations}

Note, that these numerical expressions are in agreement with those demonstrated in \cite{Zyla:2020zbs}.

Taking into account the values given above, the exact results (\ref{10}), (\ref{t2nm}-\ref{t4M}) and data from Table~\ref{Table3}, we arrive to the following expressions:
\begin{subequations}
\begin{eqnarray}
\label{McECH-method}
\frac{M_c}{1~\rm{GeV}}&\approx & 1.270+0.212+0.205+0.289+0.529 \\ \nonumber
&+&\bigg\{\underbrace{1.105+2.316}_{ECH}; ~~ \underbrace{1.044+ 2.124 }_{ECH\; direct}; ~~ \underbrace{1.160+3.303}_{FL}; \\ \nonumber
&&~~~ \underbrace{0.794+2.540 }_{FL,\; M\rightarrow \overline{m}}; ~~ \underbrace{1.316+4.011 }_{r-n} \bigg\}, \\
\label{MbECH-method}
\frac{M_b}{1~\rm{GeV}}&\approx & 4.180+0.398+0.198+0.144+0.135 \\ \nonumber
&+&\bigg\{\underbrace{0.135+0.135}_{ECH}; ~~ \underbrace{0.137+ 0.143 }_{ECH\; direct}; ~~ \underbrace{0.171+0.258}_{FL}; \\ \nonumber
&&~~~ \underbrace{0.117+0.197 }_{FL,\; M\rightarrow \overline{m}}; ~~ \underbrace{0.176+0.283 }_{r-n} \bigg\}, \\
\label{MtECH-method}
\frac{M_t}{1~\rm{GeV}}&\approx & 164.300+7.552+1.608+0.496+0.195  \\ \nonumber
&+&\bigg\{\underbrace{0.073+0.025}_{ECH}; ~~ \underbrace{0.083+ 0.037 }_{ECH\; direct}; ~~ \underbrace{0.126+0.084}_{FL}; \\ \nonumber
&&~~~ \underbrace{0.086+0.064 }_{FL,\; M\rightarrow \overline{m}}; ~~ \underbrace{0.112+0.079 }_{r-n} \bigg\}.
\end{eqnarray}
\end{subequations}

The terms in braces are the numerical contributions of the fifth and sixth orders gotten within the various estimate approaches considered by us in this work. Despite the fact that these  values  are approximate, they reflect the specific behavior of the high-order PT corrections to the relation between the
pole and $\rm{\overline{MS}}$-scheme running masses of heavy quarks, viz
\begin{itemize}
\item 
For the case of the $c$-quark
the numerical PT QCD corrections in (\ref{McECH-method}) (which  are starting to 
increase from the   $\mathcal{O}(a^3_s)$ level) are keeping on their asymptotic growth at the $\mathcal{O}(a^5_s)$ and 
$\mathcal{O}(a^6_s)$ orders. Indeed, the five-loop contribution is almost 2 times larger than the four-loop expression and the six-loop correction is 
 more than 2 times greater than the five-loop one being estimated and is even larger than the 
first term of this PT series. This effect is strongly related to both the influence of the moderately large value of the coupling constant $\alpha_s(\overline{m}^2_c)$ and to the renormalon contributions to the ratio $M_c/\overline{m}_c(\overline{m}^2_c)$. In this regard, in the modern high-precisely phenomeno- logically-oriented studies it is more appropriate to use the concept of the running $c$-quark mass that does not suffer from the factorial renormalon behavior. 
\item 
The estimates made for the $b$-quark signals that the asymptotic nature of Eq.(\ref{MbECH-method}) is starting to reveal itself from the $\mathcal{O}(a^4_s)$-contribution (except for the \textit{FL}, $M\rightarrow \overline{m}$ results where the $\mathcal{O}(a^5_s)$ contribution is less than the $\mathcal{O}(a^4_s)$ one). Note also that in the case of the application of the ECH-approach we observe the peculiar stabilization feature of the four-, five- and six-loop corrections to the on-shell-$\rm{\overline{MS}}$ mass: within the accuracy considered by us in (\ref{MbECH-method}) they coincide. However, due to the existing theoretical uncertainties discussed in Sec.~\ref{sec4} we do not consider this observation as a physical one.
\item 
The relation (\ref{MtECH-method}) demonstrates a decrease of the $\mathcal{O}(a^5_s)$ and $\mathcal{O}(a^6_s)$-corrections in all estimated approaches being investigated by us. 
Therefore,  the asymptotic behavior of the PT series for the ratio $M_t/\overline{m}_t(\overline{m}_t^2)$ is not yet manifesting itself at the six-loop level. 
\end{itemize}

One should emphasize that in fact the coefficients $t^M_k$  depends substantially on a choice of the scale parameter $\mu$ (see Eqs.(\ref{expansion}-\ref{t^M}) and \cite{Beneke:1994sw, Beneke:1998ui, Pineda:2001zq, Beneke:2016cbu} as well). For instance, shifting it from $\mu=\overline{m}_c$ to $\mu=3\; {\rm{GeV}}$ one can delay the order of manifestation of the renormalon factorial 
growth in corrections to the ratio $M_c/\overline{m}_c(\mu^2)$ (see e.g. \cite{Chetyrkin:2009fv, Chetyrkin:2010ic, Marquard:2016dcn, Dehnadi:2011gc, Alekhin:2017kpj}) and move it to the fourth order of PT (as in the case of the $b$-quark).

Note also that the effects of the massive lighter quarks in the coefficients $t^M_k$ are no less essential than the RG-controll- able ones responsible for the shift of the renormalization scale. They are rather important in both theoretical and phenomenological studies related to the determination of the charm, bottom and top-quark masses (see e.g. \cite{Ayala:2014yxa, Ayala:2016sdn, Dehnadi:2011gc, Mateu:2017hlz, Hoang:2017btd}). These massive effects were exactly calculated in \cite{Gray:1990yh, Bekavac:2007tk} (see also the recent work \cite{Fael:2020bgs}). However, in this paper we do not study the extra theoretical uncertainties related to the incorporation in analysis of the effects of massive lighter quarks.

Using the known results of \cite{Ball:1995ni} one may analyze the asymptotic structure of the relation between the pole and  running $t$-quark masses in more details. Combining the six-loop \textit{FL}-expression from (\ref{MtECH-method}) with the results of \cite{Ball:1995ni} normalized at $\mu^2=\overline{m}^2_q$ and utilizing the NNA approximation, one can arrive to the 
following numerical representation for the top-quark pole mass:
\begin{eqnarray}
\label{Mt-FL-9}
\frac{M^{FL}_t}{1~\rm{GeV}}&\approx  & 164.300+7.552+1.608+0.496+0.195 \\\nonumber
&+&\boxed{0.126+0.084+0.066+0.062+0.065 \dots}
\end{eqnarray}

This approximate expression indicates that in the case of the top quark the first traces of the asymptotic nature of the corresponding perturbative QCD series is observed above the seventh order of PT. Indeed, the contributions of the seventh, eighth and ninth orders are comparable to each other. Further, using the higher-order $t$-quark estimates for the coefficients $t^M_k$, obtained in \cite{Hoang:2017suc} with help of the IRR-based formula (\ref{renormalon-dom}), we conclude that in this case the numerical contributions to the relation between $M_t$ and $\overline{m}_t(\overline{m}^2_t)$ 
 are close to the FL-ones presented in (\ref{Mt-FL-9}).
  Therefore, the statement about the manifestation of the asymptotic behavior of the PT series for the ratio $M_t/\overline{m}_t(\overline{m}^2_t)$ after the seventh order seems to us quite reliable.

\section{Conclusion}

In this work we have estimated the values of the $\mathcal{O}(a^5_s)$ and $\mathcal{O}(a^6_s)$-contributions to the relation between the pole and running masses of heavy quarks. For this aim we have utilized three different approaches, namely the effective charges motivated method in its two variants (with and without the $\pi^2$-effects of the analytic continuation from the Euclidean to Minkowskian region), the Naive-Nonabelianization procedure applied to the results of calculations of the leading renormalon-type terms 
and the technique based on the application of the renormalon asymptotic formula with the normalization factor fixed in two ways. The ``kinematic'' analytical continuation effects, which were modeled with help of the  K\"allen-Lehmann type dispersion relation, have been associated by us with the $\pi^{2n}$-terms typical to the Minkowskian on-shell scheme. 

As a result of these estimates we have obtained that at the fixed number of massless quarks the approximate values of the
coefficients $t^M_5$ and $t^M_6$ evaluated by all three approaches are qualitatively consistent with each other (on average with a factor two). Further, using these results we have studied the flavor dependencies of these terms and  established their sign-alternating character in $n_l$ (as in the case of the already exactly calculated two, three and four-loop ones). Herewith, we especially emphasize that upon studying of the $n_l$-structure of the coefficients $t^M_5$ and $t^M_6$ estimated with help of 
the renormalon asymptotic formula the detailed information on the normalization factor $N_m$ in the expression (\ref{renormalon-dom}) plays the essential role.

Further we have considered the asymptotic structure of the relation between the pole $M_q$ and $\rm{\overline{MS}}$-scheme running masses $\overline{m}_q(\overline{m}^2_q)$ of the real heavy quarks. In comparison with the cases of the charm and bottom quarks where the asymptotic behavior manifests itself in the third and fourth orders of PT correspondingly,  the asymptotic nature of the analogous PT series for the top quark seems to reveal itself above the seventh order. Therefore, the concept of the top-quark pole mass may be used safely in the modern phenomenologically and theoretically oriented studies.

\begin{acknowledgements}
We would like to thank V.M. Braun, K.G. Chetyr- kin, L. Dudko, A.G. Grozin, M. Mangano, V.  Mateu, D.G.C. Mckeon, S.V. Mikhailov, S. Moch, P. Nason, A.F. Pikelner and A. Pineda for useful comments and fruitful discussions at different stages of the studies described in this manuscript. One of us (ALK) is grateful to colleagues from CERN-TH for inviting him to present a seminar (15.11.19) based on the definite results of this work and for valuable remarks, which were also taken into account. The work of VSM was supported by the Foundation for the Advancement of Theoretical Physics and Mathematics ``BASIS'', 
grant No. 19-1-5-114-1.
\end{acknowledgements}

\appendix

\section{Application of the least squares method}\label{A}

Let us discuss in more details the features of application of the least squares method. Following the studies done in \cite{Kataev:2018sjv} we use the results of semi-analytical  calculations of the term $t^M_4$  at the fixed number of massless quarks \cite{Marquard:2016dcn} and obtain the following overdetermined system of linear equations with two unknown parameters $t^M_{4, 0}$ and  $t^M_{4, 1}$ normalized at the point $\mu^2=\overline{m}^2_q$ and defined in (\ref{zm4}):
\begin{equation}
\label{system}
\begin{pmatrix}
    1 & 3 \\ 1 & 4 \\ 1 & 5  \\ 1 & 6  \\ 1 & 7 \\ 1 & 8 \\ 1 & 9 \\ 1 & 10 \\ 1 & 11 \\ 1 & 12 \\ 1 & 13 \\ 1 & 14 \\ 1 & 15 \\ 
   \end{pmatrix}
   \begin{pmatrix}
 \\  t^M_{4, 0} \\ \\
   \\  t^M_{4, 1} \\ \\
   \end{pmatrix}
   = \begin{pmatrix} 
   1330.44\pm 1.74 \\
   584.72\pm 1.77 \\
   -160.99\pm 1.80 \\
   -906.72\pm 1.84 \\
   -1652.44\pm 1.87\\
   -2398.16\pm 1.91\\
   -3143.88\pm 1.94\\
   -3889.61\pm 1.98\\
   -4635.32\pm 2.01\\
   -5381.04\pm 2.05\\
   -6126.77\pm 2.08\\
   -6872.49\pm 2.12\\
   -7618.21\pm 2.16\\
   \end{pmatrix}
   \end{equation}
   
Herewith, we restrict ourselves by the consideration of $n_l$ from the range  $3\leq n_l\leq 15$, where the lower bound is fixed  by us keeping in mind  that we analyze the behavior of perturbative series for the 
relation between the pole and running masses of {\it  heavy} quarks,  while the upper bound is following from the  Banks-Zaks ansatz $n_l<31/2$ \cite{Banks:1981nn},  which insures that in the considered region of $n_l$ the QCD asymptotic freedom  property is not violated.
   
To apply the LSM for solving  the system (\ref{system}) one should first
introduce  the  $\Phi$-function, which is equal to the sum of the squares of the deviations of all equations in this system:
\begin{equation}
\label{Phi-2unkn}
\Phi=\sum\limits_{s=1}^{N} (t^M_{4, 0}+t^M_{4, 1} n_{l_s}-y_{l_s})^2~,
\end{equation}
   where index $s$ runs through all values which are equal to the  number $N$ of equations of (\ref{system}) (in our case $N=13$), $y_{l_s}$ are the numbers presented on the r.h.s. of this system with their uncertainties $\Delta y_{l_s}$.
   
The LSM solutions of the overdetermined system (\ref{system}) correspond to the values of the terms $t^M_{4, 0}$ and  $t^M_{4, 1}$, for which the function $\Phi(t^M_{4, 0}, t^M_{4, 1})$ has a minimum, defined by the following requirements:
    \begin{equation}
    \label{conditions}
\frac{\partial\Phi}{\partial t^M_{4, 0}}=0 ~, ~~~~~~ \frac{\partial \Phi}{\partial t^M_{4, 1}}=0 ~.
\end{equation}

The requirements (\ref{conditions}) lead to the following system of two equations with two unknowns $t^M_{4, 0}$ and  $t^M_{4, 1}$:
\begin{equation}
\label{syst-LSM-2}
\left\{
\begin{aligned}
t^M_{4, 0}\sum\limits_{s=1}^{N}1 +t^M_{4, 1}\sum\limits_{s=1}^{N} n_{l_s} &= \sum\limits_{s=1}^{N} y_{l_s}~, \\
t^M_{4, 0}\sum\limits_{s=1}^{N} n_{l_s}+t^M_{4, 1} \sum\limits_{s=1}^{N} n^2_{l_s}&=\sum\limits_{s=1}^{N} n_{l_s}y_{l_s}~.
\end{aligned}
\right.
\end{equation}

The LSM  uncertainties of the solutions $t^M_{4, 0}$ and $t^M_{4, 1}$ of the system (\ref{syst-LSM-2}), related to the 
inaccurate knowledge of the terms $y_{l_s}$,
are fixed by the law of accumulation of errors:
\begingroup\makeatletter\def\f@size{9}\check@mathfonts
\def\maketag@@@#1{\hbox{\m@th\normalsize\normalfont#1}}
\begin{align}
\label{unc-40}
\Delta_{t^M_{4, 0}} &=\sqrt{\sum\limits_{s=1}^{N} \left( \frac{\partial t^M_{4, 0}}{\partial y_{l_s}} \Delta  y_{l_s} \right)^2+2\sum\limits_{i=1}^{N}\sum\limits_{j<i}\frac{\partial t^M_{4, 0}}{\partial y_{l_i}}\frac{\partial t^M_{4, 0}}{\partial y_{l_j}}\Delta  y_{l_i}\Delta  y_{l_j}
} \\ \nonumber
&=\frac{\sum\limits_{s=1}^{N} \Delta y_{l_s}\bigg(\sum\limits_{i=1}^{N} n^2_{l_i}-n_{l_s}\sum\limits_{i=1}^{N} n_{l_i}\bigg)}{N\sum\limits_{s=1}^{N} n^2_{l_s}-\bigg(\sum\limits_{s=1}^{N} n_{l_s}\bigg)^2}~, \\  
\label{unc-41}
\Delta_{t^M_{4, 1}}&=\sqrt{\sum\limits_{s=1}^{N} \left( \frac{\partial t^M_{4, 1}}{\partial y_{l_s}} \Delta  y_{l_s} \right)^2+2\sum\limits_{i=1}^{N}\sum\limits_{j<i}\frac{\partial t^M_{4, 1}}{\partial y_{l_i}}\frac{\partial t^M_{4, 1}}{\partial y_{l_j}}\Delta  y_{l_i}\Delta  y_{l_j}
} \\ \nonumber
&=\frac{\sum\limits_{s=1}^{N} \Delta y_{l_s}\bigg(N \;n_{l_s}-\sum\limits_{i=1}^{N} n_{l_i}\bigg)}{N\sum\limits_{s=1}^{N} n^2_{l_s}-\bigg(\sum\limits_{s=1}^{N} n_{l_s}\bigg)^2}~.
\end{align}
\endgroup

In these expressions the second term under the square root reflects the effect of the correlation between uncertainties shown in the system (\ref{system}).

However, if there were even no errors in  (\ref{system}), the LSM  would still provide uncertainties related to the quality of the reproducing of the input data. 
These uncertainties can be directly calculated from the following formulas:
\begin{eqnarray}
\label{sigma-t40}
\sigma_{t^M_{4, 0}}&=&\sqrt{\frac{\Phi(t^M_{4,0}, t^M_{4,1})}{N-2}\cdot\frac{\sum\limits_{s=1}^{N} n^2_{l_s}}{ N\sum\limits_{s=1}^{N} n^2_{l_s}-\bigg(\sum\limits_{s=1}^{N} n_{l_s}\bigg)^2}}~, \\
\label{sigma-t41}
\sigma_{t^M_{4, 1}}&=&\sqrt{\frac{\Phi(t^M_{4,0}, t^M_{4,1})}{N-2}\cdot\frac{N}{ N\sum\limits_{s=1}^{N} n^2_{l_s}-\bigg(\sum\limits_{s=1}^{N} n_{l_s}\bigg)^2}}~,
\end{eqnarray}
where $\Phi(t^M_{4,0}, t^M_{4,1})$ is the minimum of the function $\Phi$ that can be obtained from the condition (\ref{conditions}) or Eq.(\ref{syst-LSM-2}). However, these errors are much smaller (more than 100 and 30 times respectively) than the ones, given in (\ref{unc-40}-\ref{unc-41}):
\begin{equation}
\label{ll}
\sigma_{t^M_{4, 0}}\ll \Delta_{t^M_{4, 0}}, ~~~~~ \sigma_{t^M_{4, 1}}\ll \Delta_{t^M_{4, 1}}.
\end{equation}
Therefore, they do not have any noticeable effect on the final uncertainties of the coefficients $t^M_{4,0}$ and  $t^M_{4,1}$ and we do not include them in the numerical analysis.

The relations (\ref{ll}) can be explained by the fact that the following sample correlation coefficient 
\begingroup\makeatletter\def\f@size{9}\check@mathfonts
\def\maketag@@@#1{\hbox{\m@th\normalsize\normalfont#1}}
\begin{equation}
\label{correlation-r}
r=\frac{N\sum\limits_{s=1}^N n_{l_s}y_{l_s}-\sum\limits_{s=1}^N n_{l_s}\sum\limits_{s=1}^N y_{l_s} }{\sqrt{N\sum\limits_{s=1}^{N} n^2_{l_s}-\bigg(\sum\limits_{s=1}^{N} n_{l_s}\bigg)^2}\sqrt{N\sum\limits_{s=1}^{N} y^2_{l_s}-\bigg(\sum\limits_{s=1}^{N} y_{l_s}\bigg)^2 }}~,
\end{equation}
\endgroup
is very close to $r=-1$, namely $r=-0.999999999998958$. In the geometric language this means that the input data points fit perfectly the  straight line. 

Juxtaposing the solutions of the system (\ref{syst-LSM-2}) with formulas (\ref{unc-40}-\ref{unc-41}) we get the  numerical values for the constant and linearly dependent on $n_l$ contributions to the four-loop expression of  $t^M_4$  with their theoretical uncertainties\footnote{
If the impact of the correlation effects is not taken into account, the LSM-uncertainties will take the values: $\Delta_{t^M_{4, 0}}=1.34$, ~ $\Delta_{t^M_{4, 1}}=0.15$.}, viz:
\begin{subequations}
\begin{eqnarray}
\label{result}
t^M_{4, 0}(\overline{m}^2_q)&=&3567.61\pm 1.62~, \\ \nonumber t^M_{4, 1}(\overline{m}^2_q)&=&-745.720\pm 0.036~.
\end{eqnarray}

The values (\ref{result}) should be compared with the results, obtained in \cite{Marquard:2016dcn}:
\begin{eqnarray}
\label{prev}
t^M_{4, 0}(\overline{m}^2_q)&=&3567.60\pm 1.64~, \\ \nonumber
t^M_{4, 1}(\overline{m}^2_q)&=&-745.721\pm 0.040~.
\end{eqnarray}
\end{subequations}

Although the central values of the results (\ref{result}) and (\ref{prev}) were obtained by different ways, they coincide.  This is an additional argument in favor of the applicability
of the least squares method and the reliability of the results of (\ref{result}).

\section{The differences in the structure of the asymptotic series in the cases of QCD and QED: the four-loop analysis}\label{B}

Here we compare the behavior of the PT series for
the relation between the pole and running masses of the heavy quarks in QCD with the corresponding one for the charged leptons in QED at the four-loop level. This is of the definite interest because the infrared renormalons leading to the fast growth of the coefficients of the aforesaid PT series in QCD are absent in QED. However, another mechanism, not related to renormalons, for investigation of the large order behavior of the perturbative series in the various quantum field models (including QED)
has been studied in a number of works (see e.g. \cite{Lipatov:1976ny, Itzykson:1977mf, Bogomolny:1978ft, Bogomolny:1981qv} and reviews \cite{ZinnJustin:1980uk, Kazakov:1980rd}). This approach, based on the technique of the expansion of the functional integral representation for the different Green functions at a non-trivial saddle points, also indicates the factorial growth of the higher order coefficients of the related PT series. Note that the results of the specific QED studies of \cite{Itzykson:1977mf, Bogomolny:1978ft}
have pointed out the sign-alternating behavior of these coefficients in the large orders $k$. In this regard, in order to consider the possible differences in the structure of the perturbative series in QCD and QED we will focus our attention on the comparing of the behavior of the PT series 
for the relation between the pole and running masses of the heavy quarks in QCD with the corresponding one for the charged leptons in QED at the four-loop level in details.

Using Eqs.(\ref{10}) and (\ref{t2nm}-\ref{t4M}) (see \cite{Marquard:2016dcn, Kataev:2018sjv}) one can get the following numerical perturbative expression for the pole and running masses of the heavy quarks:
\begin{eqnarray}
\label{p-r-new}
M_q&\approx &\overline{m}_q(\overline{m}^2_q )(1+1.3333\overline{a}_s+(-1.0414n_l \\ \nonumber
&+&13.443)\overline{a}^2_s+(0.6527n^2_l-26.655n_l+190.60)\overline{a}^3_s \\ \nonumber
&+&(-0.6781n^3_l+43.396n^2_l+(-745.720\pm 0.036)n_l \\ \nonumber
&+&3567.61\pm 1.62)\overline{a}^4_s )~,
\end{eqnarray}
where $\overline{a}_s=\alpha_s(\overline{m}^2_q)/\pi$. Note that the presented uncertainties of the four-loop terms are very small and do not affect the \textit{asymptotic} behavior of the ratio $M_q/\overline{m}_q(\overline{m}^2_q)$. Therefore, in principle, they can be omitted in studies of this Appendix.

For the specific cases of the charm, bottom and top quarks ($n_l=3,4,5$ respectively) the expression (\ref{p-r-new}) leads to the following relations:
\begin{subequations}
\begin{eqnarray}
\label{Mcharm}
M_c&\approx &\overline{m}_c(\overline{m}^2_c )( 1+1.333\;\overline{a}_s+10.318\;\overline{a}^2_s+116.49\;\overline{a}^3_s \\ \nonumber
&+& (1702.70\pm 1.62)\;\overline{a}^4_s), \\ 
\label{Mbottom}
M_b&\approx &\overline{m}_b(\overline{m}^2_b)( 1+1.3333\;\overline{a}_s+9.277\;\overline{a}^2_s+94.41\;\overline{a}^3_s \\ \nonumber
&+&(1235.66\pm 1.63)\;\overline{a}^4_s), \\ 
\label{Mtop}
M_t&\approx &\overline{m}_t(\overline{m}^2_t)( 1+1.3333\;\overline{a}_s+8.236\;\overline{a}^2_s+73.63\;\overline{a}^3_s \\ \nonumber
&+&(839.14\pm 1.63)\;\overline{a}^4_s),
\end{eqnarray}
\end{subequations}
where the uncertainties of the four-loop terms are the mean-square errors following from (\ref{p-r-new}). 

The expressions (\ref{Mcharm}-\ref{Mtop}) demonstrate the asymptotic character of the corresponding perturbative QCD series. Indeed, all relations contain significantly growing and strictly sign-constant coefficients.   

Turn now to the study of the PT series for the relation between the pole $M_l$ and running masses $\overline{m}_l(\mu^2)$ of the charged leptons ($e, \mu, \tau$) in QED. In the case of the electron and muon their pole masses are 
the directly measurable parameters. In spite of the fact that the heavy $\tau$-lepton is decaying rather fast, one can also introduce as its main characteristic the pole mass as well. It may be extracted, for instance, from the corresponding experimental 
data for the threshold behavior of the $\tau^+\tau^-$ total cross-section production in $e^+e^-$ collisions 
(see e.g. \cite{Anashin:2007zz}). However, like in the case of quarks, it is also possible to define the  $\rm{\overline{MS}}$-scheme 
running masses of the charged leptons which may be also used in the analysis of the experimental data \cite{Xing:2007fb}.
Let us consider the structure of the ratio $M_l/\overline{m}_l(\overline{m}^2_l)$ in more details.

Using the $U(1)$-limit of the results of the diagram calculations \cite{Tarrach:1980up, Gray:1990yh, Fleischer:1998dw, Melnikov:2000qh, Chetyrkin:1999qi, Lee:2013sx} and \cite{Marquard:2016dcn} of the coefficients $t^M_k$ performed within the $SU(N_c)$ theory with their decomposition into the Casimir operators (or the recent four-loop results of the explicit numerical computations \cite{Laporta:2020fog}), it is possible to get the on-shell-$\rm{\overline{MS}}$ mass relation for the charged leptons in QED in the $\mathcal{O}(a^4)$ approximation:
\begin{eqnarray}
\label{tM4-QED}
M_l&\approx & \overline{m}^2_l(\overline{m}^2_l)(1+\overline{a}+(-1.56205N_l+0.1659)\overline{a}^2 \\ \nonumber
&+&(1.95808N^2_l-0.4726N_l-2.131)\overline{a}^3 \\ \nonumber
&+&(-4.06885N^3_l+2.1637N^2_l+(-2.415\pm 0.178)N_l \\ \nonumber
&+& \bigg\{ 7.49\pm 1.03; ~~ 7.4727\bigg\})\overline{a}^4),
\end{eqnarray}
where $\overline{a}=\alpha(\overline{m}^2_l)/\pi$ is the QED coupling constant defined in the $\rm{\overline{MS}}$-scheme and $N_l$ is the number of the massless charged leptons.
The first four-loop $N_l$-independent (the $N_l$-dependent one as well) term in the curly braces corresponds to the abelian $U(1)$-limit of the results of the semi-analytical calculations \cite{Marquard:2016dcn} carried out for the case of the $SU(N_c)$ theory with $n_l$ massless quarks. The second one follows from the recent high-precision (about 1100 digits) four-loop computations of the  
on-shell mass renormalization constant $Z^{{\rm{OS}}}_m$ performed in QED in \cite{Laporta:2020fog} (see also \cite{Laporta:2018eos} where the wave function renormalization constant $Z^{\rm{OS}}_2$ was also found)
  for the case of $N_l=0$. It is worth emphasizing that the results of \cite{Laporta:2020fog} are in very good agreement with the ones following from \cite{Marquard:2016dcn}. 
  
Taking into account Eq.(\ref{tM4-QED}) and keeping in mind that for the cases of the electron, muon and $\tau$-lepton one should set  $N_l=0, 1, 2$  respectively, we arrive to the following expressions:
\begin{subequations}
\begin{eqnarray}
\label{meme}
M_e&\approx&\overline{m}_e(\overline{m}^2_e)(1+\overline{a}+0.1659\overline{a}^2-2.131 \overline{a}^3 \\ \nonumber
&+& \bigg\{ 7.49\pm 1.03; ~~ 7.473\bigg\}\overline{a}^4), \\
\label{mumu}
M_{\mu}&\approx&\overline{m}_{\mu}(\overline{m}^2_{\mu})(1+\overline{a}-1.3961 \overline{a}^2-0.646 \overline{a}^3 \\ \nonumber
&+& \bigg\{ 3.17\pm 1.05; ~~ 3.153\pm 0.178\bigg\} \overline{a}^4),\\
\label{mtau}
M_{\tau}&\approx&\overline{m}_{\tau}(\overline{m}^2_{\tau})(1+\overline{a}-2.9582\overline{ a}^2+4.756 \overline{a}^3 \\ \nonumber
&+& \bigg\{ -21.24\pm 1.09; ~~ -21.253\pm 0.356\bigg\}   \overline{a}^4),
\end{eqnarray}
\end{subequations}
where the uncertainties of the four-loop terms are the mean-square errors following from (\ref{tM4-QED}).

Unlike Eqs.(\ref{Mcharm}-\ref{Mtop}) the
expressions (\ref{meme}-\ref{mtau}) demonstrate the absence of any regular sign-constant or sign-alternating structure of  the related PT series (besides the case of the $\tau$-lepton). The same feature is observed when the running $\overline{\rm{MS}}$-scheme QED parameters (the masses of charged leptons  and  coupling constant) 
are normalized at the scale $\mu^2=M^2_l$ (in this case the sign-alternating structure of the relation between $M_{\tau}$ and $\overline{m}_{\tau}(M^2_{\tau})$ is not manifested itself anymore). Therefore, the
point of view appearing from time-to-time in the literature  that the asymptotic perturbative series in the QED should have sign-alternating structure, which is based in part on the theoretical studies presented in \cite{Itzykson:1977mf,
Bogomolny:1978ft}, seems to be not a general rule. Note that the sign-alternating behavior is realized nowadays only in the perturbative expression for the anomalous magnetic moment of electron, which is known at present with high precision up to the five-loop term (for the most recent results of its numerical evaluation see \cite{Aoyama:2017uqe} and \cite{Volkov:2019phy}).

One should mention that the issue of the non-regular behavior of the corrections to the relation between the pole and running masses of the charged leptons in QED was first raised in \cite{Kataev:2009} upon the three-loop analysis of the numerical expressions for this relation being analytically evaluated later on in \cite{Baikov:2012rr}. It may be of interest to understand better the discussed non-regular structure of the asymptotic QED series in the future.


\begin{thebibliography}{130}

\bibitem{Tarrach:1980up}
  R.~Tarrach,
  Nucl.\ Phys.\ B {\bf 183} (1981) 384.

\bibitem{Gray:1990yh}
N.~Gray, D.~J.~Broadhurst, W.~Grafe and K.~Schilcher,
Z. Phys. C \textbf{48} (1990), 673-680

\bibitem{Avdeev:1997sz}
  L.~V.~Avdeev and M.~Y.~Kalmykov,
  Nucl.\ Phys.\ B {\bf 502} (1997) 419 

\bibitem{Fleischer:1998dw} 
  J.~Fleischer, F.~Jegerlehner, O.~V.~Tarasov and O.~L.~Veretin,
  Nucl.\ Phys.\ B {\bf 539} (1999) 671.
  Erratum: [Nucl.\ Phys.\ B {\bf 571} (2000)  511]
  
\bibitem{Melnikov:2000qh} 
  K.~Melnikov and T.~v.~Ritbergen,
  Phys.\ Lett.\ B {\bf 482} (2000) 99
  
\bibitem{Chetyrkin:1999qi} 
  K.~G.~Chetyrkin and M.~Steinhauser,
  Nucl.\ Phys.\ B {\bf 573} (2000) 617
  
\bibitem{Lee:2013sx} 
  R.~Lee, P.~Marquard, A.~V.~Smirnov, V.~A.~Smirnov and M.~Steinhauser,
  JHEP {\bf 1303} (2013) 162 
  
\bibitem{Ball:1995ni} 
  P.~Ball, M.~Beneke and V.~M.~Braun,
  Nucl.\ Phys.\ B {\bf 452}, 563 (1995)
  
\bibitem{Marquard:2015qpa} 
  P.~Marquard, A.~V.~Smirnov, V.~A.~Smirnov and M.~Steinhauser,
  Phys.\ Rev.\ Lett.\  {\bf 114}, (2015)  no. 14, 142002
  
\bibitem{Kataev:2015gvt}
  A.~L.~Kataev and V.~S.~Molokoedov,
  Eur.\ Phys.\ J.\ Plus {\bf 131} (2016) no.8,  271
  
\bibitem{Kiyo:2015ooa}
  Y.~Kiyo, G.~Mishima and Y.~Sumino,
  JHEP {\bf 1511} (2015) 084
  
\bibitem{Beneke:1994qe} 
  M.~Beneke and V.~M.~Braun,
  Phys.\ Lett.\ B {\bf 348}, (1995) 513 

\bibitem{Bigi:1994em}
  I.~I.~Y.~Bigi, M.~A.~Shifman, N.~G.~Uraltsev and A.~I.~Vainshtein,
  Phys.\ Rev.\ D {\bf 50} (1994) 2234
  
\bibitem{Beneke:1994sw} 
  M.~Beneke and V.~M.~Braun,
  Nucl.\ Phys.\ B {\bf 426},  (1994)  301
  
\bibitem{Beneke:1994rs} 
  M.~Beneke,
  Phys.\ Lett.\ B {\bf 344}, (1995)  341 
 
\bibitem{Beneke:1998ui} 
  M.~Beneke,
  Phys.\ Rept.\  {\bf 317}, (1999) 1 
 
\bibitem{Marquard:2016dcn} 
  P.~Marquard, A.~V.~Smirnov, V.~A.~Smirnov, M.~Steinhauser and D.~Wellmann,
  Phys.\ Rev.\ D {\bf 94},  (2016) no. 7, 074025 
  
\bibitem{Kataev:2018sjv}
  A.~L.~Kataev and V.~S.~Molokoedov,
  Theor.\ Math.\ Phys.\  {\bf 200} (2019) no.3,  1374
 
\bibitem{Dyson:1952tj}
F.~J.~Dyson,
Phys. Rev. \textbf{85} (1952), 631-632

\bibitem{Chetyrkin:2009fv}
K.~G.~Chetyrkin, J.~H.~Kuhn, A.~Maier, P.~Maierhofer, P.~Marquard, M.~Steinhauser and C.~Sturm,
Phys. Rev. D \textbf{80} (2009), 074010

\bibitem{Chetyrkin:2010ic}
K.~Chetyrkin, J.~H.~Kuhn, A.~Maier, P.~Maierhofer, P.~Marquard, M.~Steinhauser and C.~Sturm,
Theor. Math. Phys. \textbf{170}, 217-228 (2012)

\bibitem{Dehnadi:2011gc}
B.~Dehnadi, A.~H.~Hoang, V.~Mateu and S.~M.~Zebarjad,
JHEP \textbf{09} (2013), 103

\bibitem{Kiyo:2015ufa} 
  Y.~Kiyo, G.~Mishima and Y.~Sumino,
  Phys.\ Lett.\ B {\bf 752}, (2016) 122 
  Erratum: [Phys.\ Lett.\ B {\bf 772}, (2017) 878]
  
\bibitem{Alekhin:2017kpj}
  S.~Alekhin, J.~Blumlein, S.~Moch and R.~Placakyte,
  Phys.\ Rev.\ D {\bf 96} (2017) no.1,  014011
  
\bibitem{Mateu:2017hlz}
  V.~Mateu and P.~G.~Ortega,
  JHEP {\bf 1801} (2018) 122
  
\bibitem{Penin:2014zaa}
  A.~A.~Penin and N.~Zerf,
  JHEP {\bf 1404} (2014) 120

\bibitem{Ayala:2014yxa}
  C.~Ayala, G.~Cvetic and A.~Pineda,
  JHEP {\bf 1409} (2014) 045

\bibitem{Ayala:2016sdn} 
  C.~Ayala, G.~Cvetic and A.~Pineda,
  J.\ Phys.\ Conf.\ Ser.\  {\bf 762},  (2016)  no. 1, 012063 
   
\bibitem{Dehnadi:2015fra}
B.~Dehnadi, A.~H.~Hoang and V.~Mateu,
JHEP \textbf{08} (2015), 155
  
\bibitem{Beneke:2014pta} 
  M.~Beneke, A.~Maier, J.~Piclum and T.~Rauh,
  Nucl.\ Phys.\ B {\bf 891}, (2015) 42 
  
\bibitem{Bazavov:2018omf}
  A.~Bazavov {\it et al.} [Fermilab Lattice and MILC and TUMQCD Collaborations],
  Phys.\ Rev.\ D {\bf 98} (2018) no.5,  054517
  
\bibitem{Butenschoen:2016lpz}
M.~Butenschoen, B.~Dehnadi, A.~H.~Hoang, V.~Mateu, M.~Preisser and I.~W.~Stewart,
Phys. Rev. Lett. \textbf{117} (2016) no.23, 232001
  
\bibitem{Corcella:2019tgt}
  G.~Corcella,
  Front.\ in Phys.\  {\bf 7} (2019) 54
  
\bibitem{Hoang:2020iah}
A.~H.~Hoang,
 Annu.\ Rev.\ Nucl.\ Part.\
Sci.\ {\bf 70} (2020) 
  
\bibitem{Zyla:2020zbs}
P.~A.~Zyla \textit{et al.} [Particle Data Group],
PTEP \textbf{2020}, (2020) no.8, 083C01 
  
  

\bibitem{Aad:2019mkw}
  G.~Aad {\it et al.} [ATLAS Collaboration],
  arXiv:1905.02302 [hep-ex].

\bibitem{Baskakov:2017jhb}
  A.~Baskakov, E.~Boos and L.~Dudko,
  EPJ Web Conf.\  {\bf 158} (2017) 04007.

\bibitem{Nason:2017cxd}
  P.~Nason,
 in {\it ``From My Vast Repertoire...,'' 
  Guido Altarelli's Legacy,} eds. S.~Forte, A.~Levy and G.~Ridolfi, World Scientific (2019) p. 123-151
  arXiv:1712.02796 [hep-ph].

\bibitem{Alekhin:2016jjz}
  S.~Alekhin, S.~Moch and S.~Thier,
  Phys.\ Lett.\ B {\bf 763} (2016) 341
  
\bibitem{Catani:2020tko}
S.~Catani, S.~Devoto, M.~Grazzini, S.~Kallweit and J.~Mazzitelli,
JHEP \textbf{08}, no.08, 027 (2020)
  
\bibitem{Beneke:2016cbu} 
  M.~Beneke, P.~Marquard, P.~Nason and M.~Steinhauser,
  Phys.\ Lett.\ B {\bf 775}, (2017) 63 
  
\bibitem{Kataev:2018mob}
  A.~L.~Kataev and V.~S.~Molokoedov,
  JETP Lett.\  {\bf 108} (2018) no.12,  777
  
\bibitem{Kataev:2018fvx}
A.~L.~Kataev and V.~S.~Molokoedov,
EPJ Web Conf. \textbf{191}, (2018) 04005 

\bibitem{PhD thesis}
V.~S.~Molokoedov, 
PhD thesis (in Russian), 
\url{http://inr.ru/rus/referat/molokoed/dis.pdf}

\bibitem{Kataev:1995vh}
  A.~L.~Kataev and V.~V.~Starshenko,
  Mod.\ Phys.\ Lett.\ A {\bf 10} (1995) 235

\bibitem{Chetyrkin:1997wm} 
  K.~G.~Chetyrkin, B.~A.~Kniehl and A.~Sirlin,
  Phys.\ Lett.\ B {\bf 402},  (1997) 359

\bibitem{Grunberg:1982fw}
  G.~Grunberg,
  Phys.\ Rev.\ D {\bf 29} (1984) 2315.

\bibitem{Stevenson:1981vj} 
  P.~M.~Stevenson,
  Phys.\ Rev.\ D {\bf 23}, (1981) 2916.

\bibitem{Pineda:2001zq} 
  A.~Pineda,
  JHEP {\bf 0106}, (2001) 022 

\bibitem{Hoang:2017suc} 
  A.~H.~Hoang, A.~Jain, C.~Lepenik, V.~Mateu, M.~Preisser, I.~Scimemi and I.~W.~Stewart,
  JHEP {\bf 1804}, (2018) 003 

\bibitem{Ayala:2019hkn}
C.~Ayala, X.~Lobregat and A.~Pineda,
Phys. Rev. D \textbf{101}, no.3, 034002 (2020)
 
\bibitem{Gross:1973id} 
  D.~J.~Gross and F.~Wilczek,
  Phys.\ Rev.\ Lett.\  {\bf 30}, (1973) 1343.
  
\bibitem{Politzer:1973fx} 
  H.~D.~Politzer,
  Phys.\ Rev.\ Lett.\  {\bf 30}, (1973) 1346.
  
\bibitem{Jones:1974mm} 
  D.~R.~T.~Jones,
  Nucl.\ Phys.\ B {\bf 75},  (1974) 531.
  
\bibitem{Caswell:1974gg} 
  W.~E.~Caswell,
  Phys.\ Rev.\ Lett.\  {\bf 33}, (1974) 244.
  
\bibitem{Egorian:1978zx} 
  E.~Egorian and O.~V.~Tarasov,
  Teor.\ Mat.\ Fiz.\  {\bf 41}, (1979) 26 
  [Theor.\ Math.\ Phys.\  {\bf 41}, (1979) 863].
  
\bibitem{Tarasov:1980au} 
  O.~V.~Tarasov, A.~A.~Vladimirov and A.~Y.~Zharkov,
  Phys.\ Lett.\  {\bf 93B}, (1980) 429.
  
\bibitem{Larin:1993tp} 
  S.~A.~Larin and J.~A.~M.~Vermaseren,
  Phys.\ Lett.\ B {\bf 303}, (1993) 334 
  
\bibitem{vanRitbergen:1997va} 
  T.~van Ritbergen, J.~A.~M.~Vermaseren and S.~A.~Larin,
  Phys.\ Lett.\ B {\bf 400}, (1997) 379 
  
\bibitem{Czakon:2004bu} 
  M.~Czakon,
  Nucl.\ Phys.\ B {\bf 710}, (2005) 485
  
\bibitem{Baikov:2016tgj} 
  P.~A.~Baikov, K.~G.~Chetyrkin and J.~H.~K\"uhn,
  Phys.\ Rev.\ Lett.\  {\bf 118}, (2017) no. 8, 082002
  
\bibitem{Herzog:2017ohr}
  F.~Herzog, B.~Ruijl, T.~Ueda, J.~A.~M.~Vermaseren and A.~Vogt,
  JHEP {\bf 1702} (2017) 090
  
\bibitem{Luthe:2017ttg}
  T.~Luthe, A.~Maier, P.~Marquard and Y.~Schroder,
  JHEP {\bf 1710} (2017) 166

\bibitem{Nachtmann:1981zg} 
  O.~Nachtmann and W.~Wetzel,
  Nucl.\ Phys.\ B {\bf 187},  (1981) 333.

\bibitem{Tarasov:1982gk} 
O.~V.~Tarasov,
 JINR-P2-82-900, Communication of the Joint Institute for Nuclear Research, Dubna, 1982 (in Russian).
Phys. Part. Nucl. Lett. \textbf{17} (2020) no.2, 109-115
  
  \bibitem{Larinmass}
  S.~A.~Larin,
 in {\it  Proc. of the Int. School ``Particles and Cosmology'',
Baksan Neutrino Observatory of INR, 1993}, p. 216-226;
  eds.  E.N. Alekseev, V.A. Matveev, Kh.S. Nirov and V.A. Rubakov (World Scientific, Singapore, 1994).

\bibitem{Vermaseren:1997fq} 
  J.~A.~M.~Vermaseren, S.~A.~Larin and T.~van Ritbergen,
  Phys.\ Lett.\ B {\bf 405}, (1997) 327 
  
\bibitem{Chetyrkin:1997dh} 
  K.~G.~Chetyrkin,
  Phys.\ Lett.\ B {\bf 404},  (1997) 161
  
\bibitem{Baikov:2014qja} 
  P.~A.~Baikov, K.~G.~Chetyrkin and J.~H.~K\"uhn,
  JHEP {\bf 1410},  (2014) 076
  
\bibitem{Luthe:2016xec}
  T.~Luthe, A.~Maier, P.~Marquard and Y.~Schröder,
  JHEP {\bf 1701} (2017) 081

\bibitem{Baikov:2018wgs}
  P.~A.~Baikov and K.~G.~Chetyrkin,
  JHEP {\bf 1806} (2018) 141

\bibitem{Laporta:2020fog}
S.~Laporta,
Phys. Lett. B \textbf{802}, 135264 (2020)

\bibitem{Radyushkin:1982kg}
  A.~V.~Radyushkin,
  JINR-E2-82-155, Communication of the Joint Institute for Nuclear Research, Dubna, 1982; ~
 JINR Rapid Commun.\  {\bf 78} (1996) 96

\bibitem{Gorishnii:1983cu}
  S.~G.~Gorishniy, A.~L.~Kataev and S.~A.~Larin,
  Sov.\ J.\ Nucl.\ Phys.\  {\bf 40} (1984) 329
   [Yad.\ Fiz.\  {\bf 40} (1984) 517].

\bibitem{Pivovarov:1991bi} 
  A.~A.~Pivovarov,
  Nuovo Cim.\ A {\bf 105}, 813 (1992).

\bibitem{LeDiberder:1992jjr}
  F.~Le Diberder and A.~Pich,
  Phys.\ Lett.\ B {\bf 286} (1992) 147.

\bibitem{Altarelli:1994vz}
  G.~Altarelli, P.~Nason and G.~Ridolfi,
  Z.\ Phys.\ C {\bf 68} (1995) 257

\bibitem{Bakulev:2010gm}
  A.~P.~Bakulev, S.~V.~Mikhailov and N.~G.~Stefanis,
  JHEP {\bf 1006} (2010) 085
  
\bibitem{Nesterenko:2017wpb}
  A.~V.~Nesterenko,
  Eur.\ Phys.\ J.\ C {\bf 77} (2017) no.12,  844

\bibitem{Broadhurst:2000yc}
D.~J.~Broadhurst, A.~L.~Kataev and C.~J.~Maxwell,
Nucl. Phys. B \textbf{592} (2001), 247-293

\bibitem{Kataev:2010zh} 
  A.~L.~Kataev and V.~T.~Kim,
  Phys.\ Part.\ Nucl.\  {\bf 41}, (2010) 946 

\bibitem{Pivovarov:2001xj}
A.~A.~Pivovarov,
Phys. Atom. Nucl. \textbf{66} (2003), 724-736

\bibitem{Bjorken:1989xw}
  J.~D.~Bjorken,
  SLAC-PUB-5103.
  
\bibitem{Kataev:2015yha} 
  A.~L.~Kataev and V.~S.~Molokoedov,
  Phys.\ Rev.\ D {\bf 92}, (2015) no. 5, 054008 
  
\bibitem{Komijani:2017vep} 
  J.~Komijani,
  JHEP {\bf 1708}, (2017) 062 
   
\bibitem{Beneke:1998rk} 
  M.~Beneke,
  Phys.\ Lett.\ B {\bf 434}, (1998) 115 
   
\bibitem{Hoang:2008yj}
A.~H.~Hoang, A.~Jain, I.~Scimemi and I.~W.~Stewart,
Phys. Rev. Lett. \textbf{101} (2008), 151602

\bibitem{Campanario:2003ix}
  F.~Campanario, A.~G.~Grozin and T.~Mannel,
  Nucl.\ Phys.\ B {\bf 663} (2003) 280
   Erratum: [Nucl.\ Phys.\ B {\bf 670} (2003) 331]
   
\bibitem{Hoang:2017btd}
  A.~H.~Hoang, C.~Lepenik and M.~Preisser,
  JHEP {\bf 1709} (2017) 099

\bibitem{Pineda:2017uby}
A.~Pineda,
[arXiv:1704.05095 [hep-ph]].

\bibitem{Ball:2014uwa} 
  R.~D.~Ball {\it et al.} [NNPDF Collaboration],
  JHEP {\bf 1504}, (2015) 040 
   
\bibitem{Kniehl:2006bg}
  B.~A.~Kniehl, A.~V.~Kotikov, A.~I.~Onishchenko and O.~L.~Veretin,
  Phys.\ Rev.\ Lett.\  {\bf 97} (2006) 042001
   
\bibitem{Chetyrkin:1997sg} 
  K.~G.~Chetyrkin, B.~A.~Kniehl and M.~Steinhauser,
  Phys.\ Rev.\ Lett.\  {\bf 79},  (1997) 2184
  
\bibitem{Bernreuther:1981sg}
W.~Bernreuther and W.~Wetzel,
Nucl. Phys. B \textbf{197}, 228-236 (1982)
[erratum: Nucl. Phys. B \textbf{513}, 758-758 (1998)]
  
\bibitem{Larin:1994va}
S.~A.~Larin, T.~van Ritbergen and J.~A.~M.~Vermaseren,
Nucl. Phys. B \textbf{438}, 278-306 (1995)
      
\bibitem{Chetyrkin:2005ia}
  K.~G.~Chetyrkin, J.~H.~Kuhn and C.~Sturm,
  Nucl.\ Phys.\ B {\bf 744} (2006) 121

\bibitem{Schroder:2005hy}
  Y.~Schroder and M.~Steinhauser,
  JHEP {\bf 0601} (2006) 051

\bibitem{Bekavac:2007tk}
S.~Bekavac, A.~Grozin, D.~Seidel and M.~Steinhauser,
JHEP \textbf{10} (2007), 006

\bibitem{Fael:2020bgs}
M.~Fael, K.~Schönwald and M.~Steinhauser,
[arXiv:2008.01102 [hep-ph]].
  
\bibitem{Banks:1981nn}
  T.~Banks and A.~Zaks,
  Nucl.\ Phys.\ B {\bf 196} (1982) 189.
  
\bibitem{Lipatov:1976ny}
L.~N.~Lipatov,
Sov. Phys. JETP \textbf{45}, (1977) 216-223 
  
\bibitem{Itzykson:1977mf}
  C.~Itzykson, G.~Parisi and J.-B.~Zuber,
  Phys.\ Rev.\ D {\bf 16} (1977) 996.
  
\bibitem{Bogomolny:1978ft}
E.~B.~Bogomolny and V.~A.~Fateev,
Phys. Lett. B \textbf{76}, 210-212 (1978)
         
\bibitem{Bogomolny:1981qv}
E.~B.~Bogomolny and Y.~A.~Kubyshin,
Sov. J. Nucl. Phys. \textbf{34}, 853-858 (1981)

\bibitem{ZinnJustin:1980uk}
J.~Zinn-Justin,
Phys. Rept. \textbf{70},  (1981) 109
  
\bibitem{Kazakov:1980rd}
D.~I.~Kazakov and D.~V.~Shirkov,
Fortsch. Phys. \textbf{28}, 465-499 (1980)
  
\bibitem{Anashin:2007zz}
  V.~V.~Anashin {\it et al.},
  JETP Lett.\  {\bf 85} (2007) 347.

\bibitem{Xing:2007fb}
  Z.~z.~Xing, H.~Zhang and S.~Zhou,
  Phys.\ Rev.\ D {\bf 77} (2008) 113016
  
\bibitem{Laporta:2018eos}
  S.~Laporta,
  PoS LL {\bf 2018} (2018) 073.

\bibitem{Aoyama:2017uqe}
  T.~Aoyama, T.~Kinoshita and M.~Nio,
  Phys.\ Rev.\ D {\bf 97} (2018) no.3,  036001
 
\bibitem{Volkov:2019phy} 
  S.~Volkov,
  Phys.\ Rev.\ D {\bf 100}, no. 9, 096004 (2019)
 
\bibitem{Kataev:2009}
A.~L.~Kataev, 
 Talk at XIV Lomonosov Conference on Elementary Particle Physics, MSU, Moscow, 19-25 August, 2009  (unpublished) \url{http://nuclphys.sinp.msu.ru/conf/lpp14/210809/Kataev.pdf}

\bibitem{Baikov:2012rr}
  P.~A.~Baikov, K.~G.~Chetyrkin, J.~H.~Kuhn and C.~Sturm,
  Nucl.\ Phys.\ B {\bf 867} (2013) 182
  

\end{thebibliography}
\end{document}